\documentclass{aa}  
\usepackage{graphicx}
\usepackage{subfigure}
\usepackage{txfonts}
\usepackage{booktabs}
\usepackage{blindtext}
\usepackage[]{hyperref}
\usepackage{xcolor}

\begin{document}

  \title{The VANDELS survey: the role of ISM and galaxy physical properties on the escape of Ly$\alpha$ emission in  $z\sim 3.5$ star-forming galaxies
                             \thanks{Based on data obtained with the European 
                                       Southern Observatory Very Large Telescope, Paranal, Chile, under Large
                                       Program 194.A-2003(EK). }
                  }

           \author{F. ~Marchi\inst{1}
                   \and L.~Pentericci\inst{1}
                   \and L.~Guaita\inst{4}
                   \and M.~Talia\inst{15,2}
                   \and M.~Castellano\inst{1}
                   \and N.~Hathi\inst{5}
                   \and D.~Schaerer\inst{3}
                   \and R.~Amorin\inst{6,16}
                   \and A. C. ~Carnall\inst{11}
                    \and S.~Charlot\inst{9}
                   \and J.~Chevallard\inst{9}
                   \and F. ~Cullen\inst{11}
                   \and S. L. ~Finkelstein\inst{13}
                   \and A. ~Fontana\inst{1}
                   \and F.~Fontanot\inst{10}
                   \and B.~Garilli\inst{7}
                   \and P.~Hibon\inst{12}
                   \and A.~ M. ~Koekemoer\inst{8}
                   \and D.~Maccagni\inst{7}
                   \and R. J. ~McLure\inst{11}
                   \and C. ~Papovich\inst{14}
                   \and L. ~Pozzetti\inst{2}
                   \and A. ~Saxena\inst{1}
                   }       
                
           \institute{INAF–Osservatorio Astronomico di Roma, via Frascati 33, 00040 Monte Porzio Catone, Italy\\
                      \email{francesca.marchi@inaf.it}
                 \and
                                 INAF--Osservatorio Astronomico di Bologna, Via Gobetti 93/3 - 40129, Bologna, Italy     
                 \and
                         Geneva Observatory, University of Geneva, ch. des Maillettes 51, CH-1290 Versoix,
                         Switzerland
                 \and
                                 Nùcleo  de  Astronomìa,  Facultad  de  Ingenierìa,  Universidad  Diego  Portales,  Av.    Ejèrcito  441,  Santiago,  Chile   
                 \and        
				         Space Telescope Science Institute, 3700 San Martin Drive, Baltimore, MD 21218, USA	
				 \and 
				 Instituto de Investigatiòn Multidisciplinar en Ciencia y Tecnologìa, Universidad de La Serena, Raùl Bitràn 1305, La Serena, Chile			
				 \and 
				 INAF--IASF Milano, via Bassini 15, I--20133, Milano, Italy      
				 \and
				       Space Telescope Science Institute, 3700 San Martin Drive, Baltimore, MD 21218, USA
				  \and 
				  Sorbonne Universites, UPMC-CNRS, UMR7095, Institut d’Astrophysique de Paris, F-75014 Paris, France
				  \and 
				  INAF-Astronomical Observatory of Trieste, via G.B. Tiepolo 11, 34143 Trieste, Italy
				  \and 
				  Institute for Astronomy, University of Edinburgh, Royal Observatory, Edinburgh EH9 3HJ, UK7
				  \and 
				  European Southern Observatory (Chile)
				  \and 
				  Department of Astronomy, The University of Texas at Austin, Austin, TX, 78712
				  \and 
				  Department of Physics and Astronomy, Texas A\&M University, College Station, TX 77843-4242, USA
				  \and
				  Dipartimento di Fisica e Astronomia, Universit`a di Bologna, Via Gobetti 93/2, I-40129, Bologna, Italy 
				  \and 
				  Departamento de F\'isica y Astronom\'ia, Universidad de La Serena, Av. Juan Cisternas 1200 Norte, La Serena, Chile. \\
		          }


   \abstract
   {}
   {We wish to investigate the physical properties of a sample of Ly$\alpha$ emitting 
   	galaxies in the VANDELS survey, with particular focus on the role of kinematics and neutral hydrogen column density 
   	in the escape and spatial distribution of Ly$\alpha$ photons.}
   {From all the Ly$\alpha$ emitting galaxies in the VANDELS Data Release 2  at $3.5\lesssim z \lesssim 4.5$, 
   	we select  a sample of 52 galaxies which also have a precise systemic redshift determination from at least one nebular emission line (HeII or CIII]). For these galaxies, we derive different physical properties (stellar mass, age, dust extinction  and star formation rate) from  spectral energy distribution (SED) fitting of the exquisite multi-wavelength photometry available in the VANDELS fields, using  a state-of-the-art spectral modeling tool, BEAGLE and the UV $\beta$ slope from  the observed photometry.
   	We characterize the Ly$\alpha$ emission in terms of kinematics, EW, FWHM and spatial extension and then estimate the velocity of the neutral outflowing gas. Thanks to the ultra-deep VANDELS spectra (up to 80 hours on-source integration) this can be achieved for individual galaxies, without relying on stacks. We then investigate the correlations between the Ly$\alpha$ properties and  the other measured properties, to study how they affect the shape and intensity of Ly$\alpha$ emission.}
   {We reproduce some of the well known correlations between Ly$\alpha$ EW and stellar mass, dust extinction and UV $\beta$ slope, in the sense that the emission line appears brighter in lower mass, less dusty and bluer galaxies. We do not find any correlation with the SED-derived star formation rate, while we find that galaxies with brighter Ly$\alpha$ tend to be more compact both in UV and in Ly$\alpha$. Our data reveal a new interesting correlation  between the  Ly$\alpha$ velocity  and the offset of the inter-stellar absorption lines with respect to the systemic redshift, in the sense that galaxies with larger inter-stellar medium (ISM) out-flow velocities show smaller Ly$\alpha$ velocity shifts. We interpret this relation  in the context  of  the shell-model scenario, where the velocity of the ISM and the HI column density contribute together in determining the Ly$\alpha$ kinematics.  In support to our interpretation, we  observe that galaxies with high HI column densities have much more extended Ly$\alpha$ spatial profiles, a sign of increased scattering. However, we do not find any evidence that the HI column density is related to any other physical properties of the galaxies, although this might be due in part to the limited range of parameters that our sample spans. }  
    {}

        \titlerunning{Ly$\alpha$ properties in VANDELS}
        \authorrunning{Marchi F.}
   \maketitle
        
\label{cap:LyACIII]}

\section{Introduction}
The Ly$\alpha$ emission line plays a fundamental role in the investigation of several astrophysical problems, from detecting very distant sources, to inferring galaxy physical parameters. Understanding the mechanisms that regulate this strong emission, is critical for interpreting the galaxy populations during the reionization epoch, since the most luminous Ly$\alpha$ emitters (LAE) are frequent sources among the population of high-z galaxies \citep[e.g.][]{stark10,pentericci09,stark+17}. In addition, it is believed that the mechanisms that regulate the escape of Ly$\alpha$  photons are also responsible for the escape of LyC photons \citep{verhamme17,steidel18,marchi18}. Constraining these processes is therefore fundamental to investigate the ionizing radiation from high redshift star-forming galaxies.

Since Ly$\alpha$ photons are mainly produced in HII regions by recombination processes, the line strength is related to the UV radiation emitted by young stars in galaxies, and so to  the on-going star formation. The escape of Ly$\alpha$ photons, depends, however, on a variety of physical processes. The interpretation from  theoretical studies is  that these photons are scattered by the neutral gas and absorbed by dust in the inter-stellar medium (ISM) \citep[e.g.][]{verhamme06,gronke16} and that these processes have a fundamental role in shaping the Ly$\alpha$ line profile, which is usually characterized by a single peak, redshifted  with respect to the systemic redshift, plus a blue-ward emission. In many LAEs, however, we can clearly distinguish two different peaks on the Ly$\alpha$ spectrum \citep[e.g.][]{yamada12,verhamme17,matthee18}, red-ward and blue-ward of the systemic position, with the red peak more commonly dominant with respect to the blue one. 
 According to models, the  shift of the red  peak, the intensity of the blue peak and its distance from the red peak critically depend on the neutral hydrogen column density (N(HI)) and the dust content of galaxies \citep[e.g.][]{verhamme15}.

 The physical properties of Ly$\alpha$ emitting galaxies and the relation between the Ly$\alpha$ emission and galaxy physical properties have been extensively investigated by several groups during the last decade, using observations of large samples of galaxies at different redshifts. In particular, a bright Ly$\alpha$ emission appears to be associated with galaxies showing  lower UV luminosities and lower metallicities \citep[e.g.][]{shapley03,reddy06} and lower dust content. It is also believed that lower stellar masses and lower SFRs characterize the brightest LAEs, even if for these two quantities a general consensus has not been yet achieved \citep[see e.g.][]{pentericci10,kornei10,hathi16,du18}. While some authors advocate a scenario in which Ly$\alpha$  emitting galaxies represent early stages in galaxy formation \citep[e.g.][]{finkelstein07,cowie11}, others find that these galaxies actually span a wide range of physical properties \citep[e.g.][]{pentericci09,finkelstein09,kornei10,finkelstein15}. \\
 
The recently completed spectroscopic survey VANDELS  \citep[a VIMOS survey of the CANDELS UDS and CDFS fields,][]{pentericci18,mclure18} has obtained spectra of star forming galaxies spanning a wide range of redshifts and with unprecedented depth. Thanks to this new data-set  we can analyze the Ly$\alpha$ line properties in  extremely deep and  moderate resolution spectra for  a large sample of $z\sim 3.5$ galaxies,  and infer  the galaxy properties that most correlate with the intensity and kinematics of this line. In particular, we  aim to assess the role of ISM kinematics and neutral hydrogen column density in the escape and distribution of Ly$\alpha$ photons. The paper is organized as follows. In Section \ref{lyasample} we describe the  criteria that we used to select the sample. In Section \ref{lyamethod} we describe the methods used to measure  the spectral properties from the deep VANDELS spectra and the physical properties from SED fitting  of the deep multi-band imaging available. In Section \ref{sec:parentvsselected} we describe how our sample relates with a more generic sample of galaxies with Ly$\alpha$ emission also from the VANDELS survey. In Section \ref{lyaresults} we present the results of this analysis and show the correlations that we found between Ly$\alpha$ and galaxies properties. In Section \ref{lyadiscussion} we finally discuss the main  result, that is the  correlation between the Ly$\alpha$ peak shift and the velocity of the gas inside the ISM, and interpret this relation in the context of the shell model,  assessing  the role of the gas kinematics and HI column density in the Ly$\alpha$ photons escape. Throughout the paper we adopt the $\Lambda$ cold dark matter ($\Lambda$-CDM) cosmological model ($H_0 = 70\, \mathrm{km\, s^{-1} Mpc^{-1}}$, $\Omega_M = 0.3$ and $\Omega_{\Lambda} = 0.7$) and express all magnitudes in the AB system.
 Since the VANDELS spectra are calibrated in air and not in vacuum, we use air wavelengths to fit the position of the different lines. We also use positive values for the EWs of emission lines and negative for absorption lines. Our EWs are always rest-frame values.

\section{Sample selection}
\label{lyasample}
To study the Ly$\alpha$ emission properties of high redshift star-forming galaxies, we exploited the extremely rich and deep VANDELS data-set. In this section we first describe the VANDELS survey and then the criteria that we applied to select our sample.
\subsection{The VANDELS survey}
VANDELS, a VIMOS survey of the CANDELS UDS and CDFS fields (PI L. Pentericci and R. McLure) is an ESO public spectroscopic survey mainly targeting high-redshift galaxies at $z>2.5$ and specifically designed to be the deepest-ever spectroscopic survey of the high-redshift Universe. The survey is presented in two companion papers,  \cite{mclure18} that provides an overview of the survey strategy  and target selection, and \cite{pentericci18} that is  focused on the observations and the first data release. VANDELS exploited the VIMOS multi-object spectrograph to obtain ultra-deep optical spectra of $\sim 2100$ high-redshift galaxies within the redshift interval $1 \lesssim z \lesssim 7$, covering a total area of $0.2 \, \mathrm{deg}^2$ centered on the CANDELS-UDS (Ultra Deep Survey) and CANDELS-CDFS (\textit{Chandra} Deep Field South) fields \citep{grogin11,koekemoer+11}. Thanks to the ultra-long exposure time strategy, each target  has been observed from a minimum of 20 hours to a maximum of 80 hours, using  the red medium-resolution grism of VIMOS, that covers a wavelength range of $4800\,\AA \leq \lambda \leq 10000\,\AA$ with an average resolution of 580. 
To select VANDELS sources, in  the CANDELS/HST areas (CDFS-HST and UDS-HST)   the photometric catalogs  based on H$_{160}$-band detections and provided by the CANDELS team \citep{galametz+13,guo+13} were adopted.  Specifically, the CDFS-HST catalog is based on 17 photometric broadband filters
\citep{guo+13}, while the UDS-HST catalog includes photometry in 19 broadband filters \citep{galametz+13}. However, since the VIMOS footprint is wider than the one of HST, many VANDELS sources fall outside the CANDELS footprints but are still covered by ground-based and Spitzer IRAC imaging. The two photometric catalogs used for the sources in the wide-field regions were generated by \cite{mclure18} using the publicly available imaging in 12 filters for UDS-GROUND and 17 filters for CDFS-GROUND,
spanning the range from U to K band in both cases. These photometric catalogs, not only will be combined with the spectroscopic data to create these unique data-set to study the high redshift galaxies properties and evolution, but have also allowed the VANDELS team to carefully pre-select the VANDELS targets on the basis of their accurate photometric redshift measures \citep{mclure18}. $\sim 85 \%$ of the galaxies were selected to be at $z \geq 3$, the majority of which being Lyman Break Galaxies (LBG), which are star-forming galaxies selected on the basis of their colors red-ward and blue-ward of the observed-frame Lyman Limit. 
The mask were prepared using the VIMOS mask preparation
software \citep{bottini05} that is distributed by ESO. The VANDELS spectroscopic data were reduced using the fully automated pipeline Easy-Life \citep{garilli12}, an
updated version of the algorithms and dataflow of the original
VIPGI system \citep{scodeggio05}, that generated wavelength- and flux-calibrated spectra directly from the raw data. 
Finally, the spectroscopic redshifts have been estimated using the Pandora software package,
within the EZ environment \citep{garilli10}.
The  same procedure adopted in VUDS was applied \citep[see][]{lefevreVIMOS}, i.e. each  spectrum was analysed by two different team members. In addition 
a final independent check by the two PIs was carried out. At present, the same VUDS quality flags \citep[see][]{lefevreVIMOS} are used to asses the spectroscopic redshift reliability.\\
The second VANDELS data release, became public at the beginning of October 2018 and contains the spectra and redshifts  of 1339 objects (557 in CDFS and 762 in UDS) with more than 200 spectra with ultra-deep 80 hours exposures \footnote{The VANDELS second data release ca be found at \href{url}{https://www.eso.org/sci/publications/announcements/sciann17139.html}}.

\subsection{The sample}

We selected all the objects in the VANDELS data release 2 which showed  Ly$\alpha$ in emission and had a secure spectroscopic redshift with VANDELS reliability flags 3 or 4 \citep[corresponding to a probability greater than 95\% for the spectroscopic redshift to be correct, see][]{pentericci18}. This sample, which we will call the \textit{parent sample}, contains 305 star-forming galaxies in the redshift range $3\lesssim z\lesssim 4.5$.  146 of these galaxies are in the CDFS field and 159 are in UDS. 154 (78 in the CDFS and 76 in UDS) fall within the CANDELS regions \citep{grogin11,koekemoer+11}, and therefore are covered by very deep multi-wavelength HST imaging, while the remaining 151 (68 and 83 in the CDFS and UDS, respectively)  are in the wide-field areas of the VANDELS fields and mainly benefit from ground-based imaging.

Since we want to study the kinematics of the  Ly$\alpha$ emission, which gives us 
important information about the ISM and the HI column density as shown by \cite{verhamme15}, we need a precise measure of the velocity  of the Ly$\alpha$ line with respect to the systemic redshift of the galaxy.
Therefore,  we restricted our  sample  to those  galaxies for which we could estimate the systemic redshift in a reliable way. The systemic redshift (z$_{sys}$) is ideally  defined  by the positions of photospheric absorption lines, that are generated by the absorption of the stellar radiation by the most external layers of the star.  The most important  photospheric absorption lines in the UV regime for highly star forming galaxies  would be the OIV $\lambda$1343$\AA$, SiII $\lambda$1417$\AA$ and SV $\lambda$1500$\AA$ \citep{talia12}. However, these lines are either too weak or even not detected in our ultra-deep spectra.
As an alternative we can use non-resonant nebular emission lines which  originate from the photoionization of nebular regions by the radiation from very massive O and B stars and can be therefore used to trace the systemic redshift.  The only nebular emission line that enters the wavelength range covered by the VANDELS spectra at redshift $\sim 4$ is the CIII] $\lambda$1909$\AA$ emission. This is a semi-forbidden line at $\lambda=1907.7 \AA$ and it is known to be a doublet at a rest-frame vacuum wavelength of $1906.68 \AA$ and  $1908.68 \AA$ (note that, since the VANDELS spectra are air-calibrated, we use in this paper the corresponding air wavelengths $1906.05 \AA$ and $1908.05 \AA$). There is however another emission line that often appears in our spectra that can trace the systemic redshift, namely  the HeII at rest-frame vacuum wavelength $1640\AA$ ($1639.83 \AA$ in air). This feature can be produced either by Wolf-Rayet stars (WR), that are thought to represent a stage in the evolution of O-type stars with masses $>20M_{\odot}$, or by young star clusters with intense star formation \citep[e.g.][]{talia12,cassata13}. Recently, \cite{schaerer19} also proposed that high-mass X-ray binaries could be the main source of nebular HeII emission in low-metallicity star-forming galaxies.

The line can show a broad profile because of the strong winds powered by the intense star formation, or can appear as a nebular narrow-line as the strong UV emission from the stars photoionizes the surrounding medium. Because of the possible
effects of the winds, this line is not always a good tracer of the systemic redshift. However the results presented in this paper do not change if we exclude the sources whose systemic redshift was derived using HeII $\lambda$1640$\AA$, suggesting that at least for our galaxies, this line provides a good estimate of the systemic redshift. 

We therefore selected from  the parent sample all those  galaxies that  showed a CIII] $\lambda$1909$\AA$ and/or HeII $\lambda$1640$\AA$ emission lines, to derive the systemic redshift. We only included objects where the CIII] $\lambda$1909$\AA$ (or HeII $\lambda$1640$\AA$) were reasonably clear of  skyline residuals and had a  S/N  $\geq 3$, to  allow a precise measurement of the line peak (see below). With these criteria we selected 60 star-forming galaxies in a redshift range of $3\lesssim z \lesssim 4.5$.

\begin{figure*}
	\centering
	\includegraphics[width=0.49\linewidth]{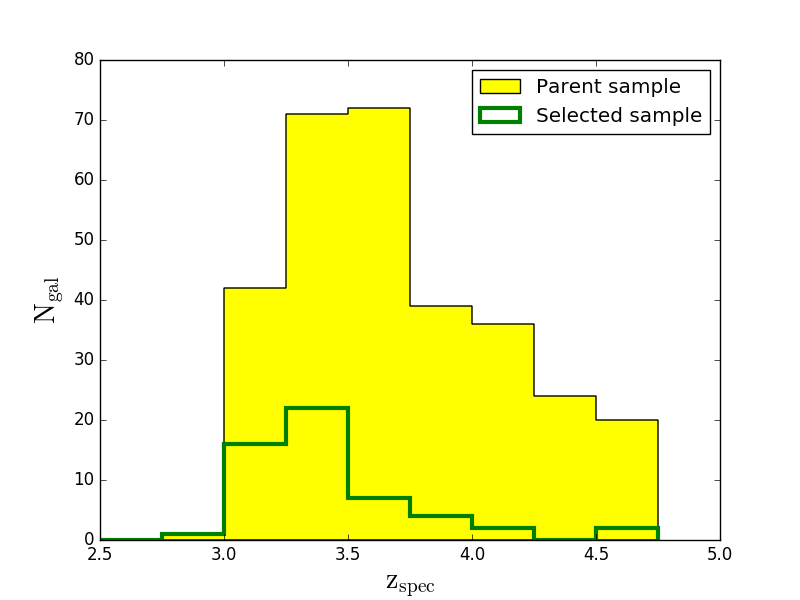}
	\includegraphics[width=0.49\linewidth]{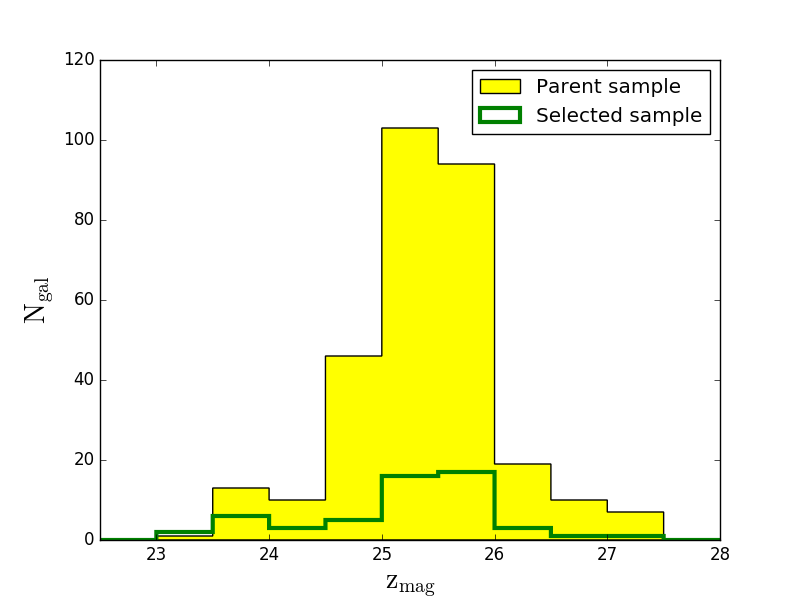}
	\caption{\small{Redshift distribution (\emph{Left panel}) and z magnitude distributions  (\emph{Right panel}) of the sources in the \textit{selected sample} (green empty histogram) and in the \textit{parent sample} (yellow histogram).}}
	\label{fig:distzezmag}
\end{figure*}

 From this sample we further excluded few objects whose Ly$\alpha$ was too noisy to give a precise estimate of the peak shift and the EW, and those where residuals from skylines fell close to the Ly$\alpha$ emission line, preventing us from  carrying out the  measures in a reliable way. We also excluded two objects with intense CIV $\lambda$1549$\AA$ emission, that could indicate the presence of an AGN.

 The final \textit{selected sample} contains 52 sources.
 29 (25) of these are in the CDFS (UDS) field.  30 (18 in the CDFS and 12 in UDS) are in the CANDELS regions \citep{grogin11,koekemoer+11}, and therefore are covered by very deep multi-wavelength HST imaging, while the remaining 24 (11 and 13 in the CDFS and UDS respectively) are in the wide-field areas of the VANDELS fields.  
Through out the paper,  we refer to this  group of galaxies as simply the  {\it selected sample}. 
 
The  distributions in redshift and z-band magnitude (that probes the rest-frame UV continuum at $\sim 1800\AA$ for all sources) of the \textit{parent sample} and the \textit{selected sample} are shown in Fig. \ref{fig:distzezmag} in yellow and green, respectively. The median redshift and the median z-band magnitude are  $\mathrm{z_{med}}=3.47$ and $\mathrm{zmag_{med}}=25.4$, for the \textit{selected sample} and $\mathrm{z_{med}}=3.47$ and $\mathrm{zmag_{med}}=25.5$, for the \textit{parent sample}. The two samples have very similar UV continuum luminosity distributions. However, since the selected sample mostly contains galaxies that also show CIII] $\lambda\, 1909\AA$ emission, we expect this sample to be in part biased to brigher Ly$\alpha$ luminosities, given that these two emissions appear to be correlated from several studies \citep[e.g.][]{shapley03,stark14,maseda17}. 

Finally, we remark that although our samples have been selected on the presence of Ly$\alpha$ emission, only a fraction could be strictly  considered LAEs if we apply the traditional threshold of $EW(Ly\alpha)>25\AA$. In our parent sample, $26\%$ of all galaxies has $EW(Ly\alpha) \geq 25 \AA$ in agreement with the statistics at this redshift \citep{shapley03,stark10}.

  
\section{Method}
\label{lyamethod}
For each galaxy in our samples, the VANDELS dataset provides not only a deep spectrum  but also deep  optical, near-IR and Spitzer photometry. To best exploit this unique legacy dataset, we therefore used the optical spectra to characterize the Ly$\alpha$ emission, both spectrally and spatially, the inter-stellar absorption lines and the CIII] $\lambda$1909$\AA$ properties,  while we  used the VANDELS photometric catalogs \citep{mclure18} to derive the physical parameters. We note that for the parent sample we only measured the Ly$\alpha$ EW from the 1-dimensional spectra and the physical properties from the available photometry.

\subsection{Spectroscopic measures}
From the VANDELS spectra we  measured the following properties.
\begin{figure*}
	\centering
	\includegraphics[width=0.8\linewidth]{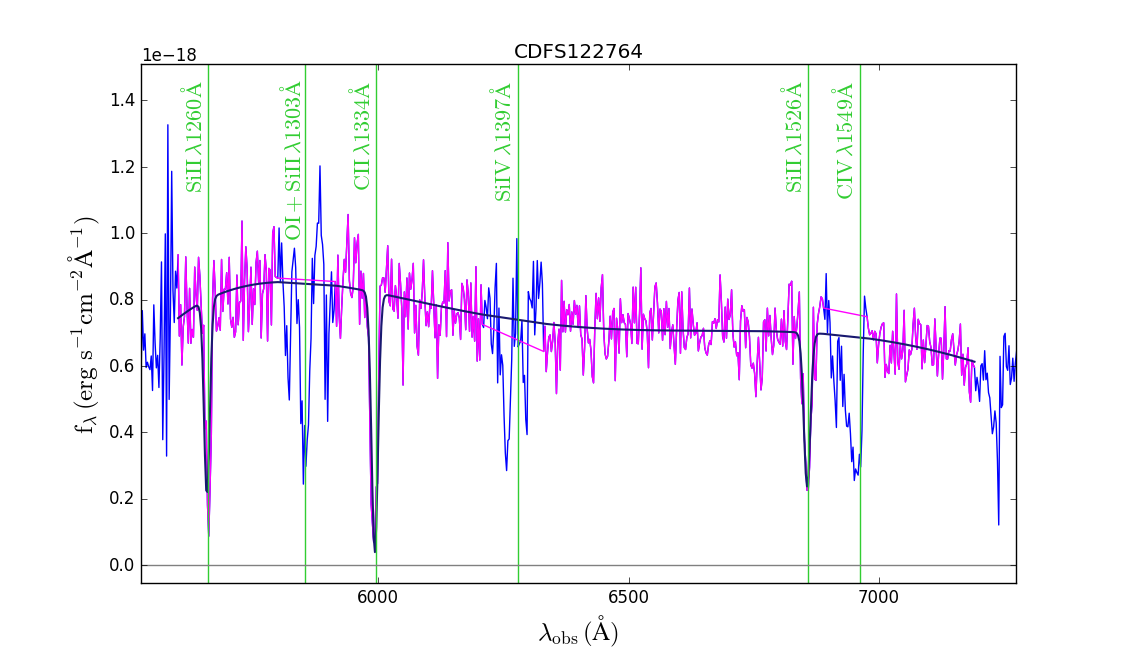}
	\includegraphics[width=0.49\linewidth]{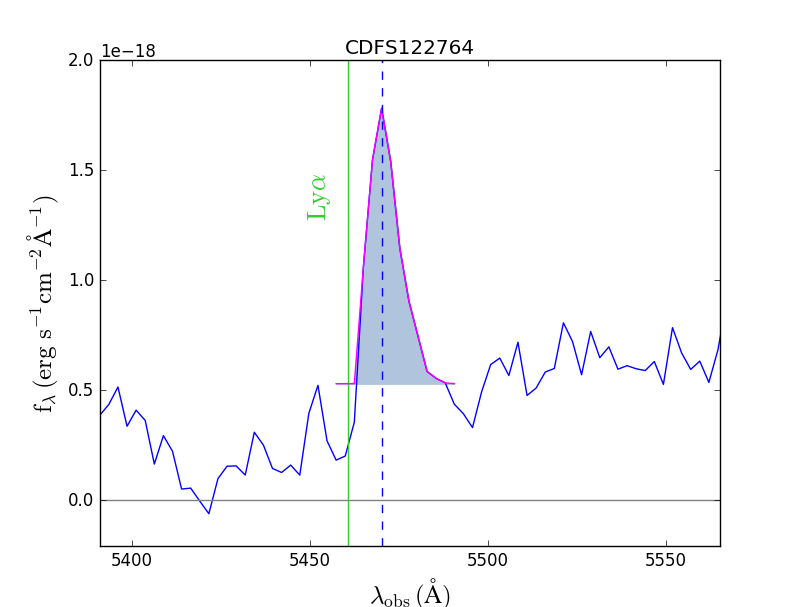}
	\includegraphics[width=0.49\linewidth]{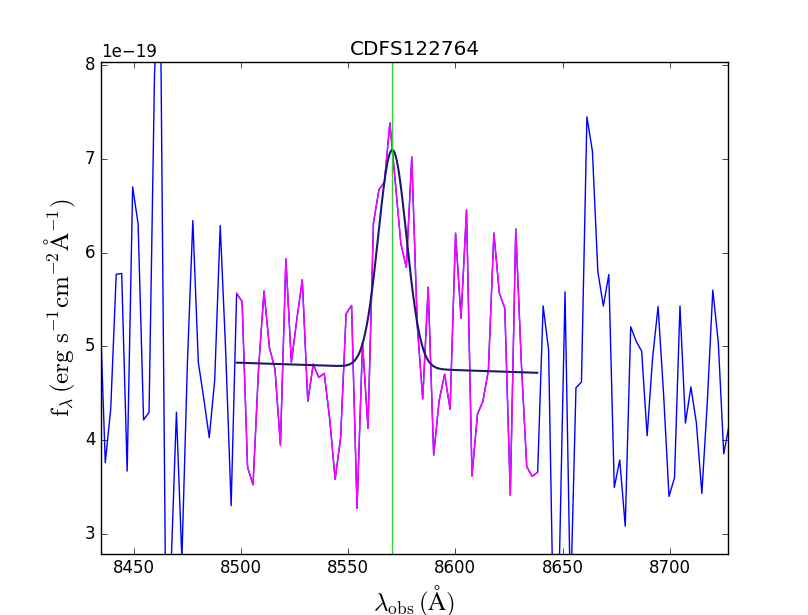}
	\caption{\small{The different panels show different regions of the spectrum of the object CDFS122764 at $z=3.49$ that we show for reference. The \textit{upper panel} shows the UV continuum between $\sim 1250-1750\, \AA$ rest-frame, where the most important absorption features are clearly visible, the \textit{lower left  panel} shows the Ly$\alpha$ line profile and the \textit{right lower panel} shows the CIII] $\lambda$1909$\AA$ emission. The magenta parts of the spectrum are  the regions that we used to perform the fits. The dark blue lines in the \textit{upper panel} and the \textit{lower right panel} are the best fits and the shaded region in the \textit{left lower panel} represent the normalized integrated flux of the Ly$\alpha$ that we used to evaluate the EW of the line. The vertical green lines show the positions of the lines at rest-frame air wavelength. The blue vertical dashed line in the \textit{left lower panel} indicates the position of the Ly$\alpha$ peak.}}
	\label{fig:examplefitspectrum}
\end{figure*}
\begin{itemize}
	\item \emph{The systemic redshift (z$_{sys}$)}
	
	We evaluated the systemic redshift from the CIII] $\lambda$1909$\AA$ line or from the HeII $\lambda$1640$\AA$ line. In the majority of galaxies with  CIII] $\lambda$1909$\AA$ emission, the two carbon lines appear blended; for these and  for the objects with HeII $\lambda$1640$\AA$ emission,  we derived the systemic redshift from the observed-frame spectra by fitting a single Gaussian profile assuming  that the continuum has a well-defined slope (linear fit).
	In the minimization procedure we took into account the VANDELS noise spectrum, that contains both the observational and instrumental uncertainties.  While  fitting the continuum, we excluded the wavelength regions with  absorption lines or where strong residuals from sky lines were present, because these regions could alter the slope of the fit. This procedure was done by visual inspection in each spectrum. We finally estimated the systemic redshift from the best-fit mean of the Gaussian profile. 	
	We show an example of a fit of the CIII] $\lambda$1909$\AA$ in the bottom right panel of Figure \ref{fig:examplefitspectrum}. The blue line is the VANDELS spectrum, the magenta range is the part of the spectrum that we are fitting and the dark blue line is the best fit. 
	For the few objects that clearly show the two peaks of the CIII] $\lambda$1909$\AA$ line due to the better S/N, we used the same procedure but including two Gaussian profiles, and we evaluated the systemic redshift directly from the fit imposing the position of the two mean peaks to be at $1906.68\,(1+z) \, \AA$ and  $1908.68\,(1+z) \, \AA$. We did not apply any restriction to the CIII] doublet ratio during the fitting procedure, but the ratios that we obtain have values in the range $\sim 1-1.5$ that are  typical values for the electronic densities of local star-forming regions \citep[see e.g.][]{CIIIdoublet,quider09}.
	To derive the errors on these measures we used the equation given by \cite{lenz92} for a Gaussian fit.
	
	\item \emph{The Ly$\alpha$ redshift and  Ly$\alpha$ velocity shift (Ly$\alpha_{shift}$)}
	
	The Ly$\alpha$ redshift was evaluated as the peak  of the Ly$\alpha$ line i.e. at the position where the flux has a maximum. Contrary to the CIII]  emission we did not attempt a Gaussian  fit of the Ly$\alpha$ line,  since its shape is known to be asymmetric  at high redshift \citep[e.g.][]{shimasaku06}.  For the few spectra (5 out of 52) where we could clearly identify  a  double peak profile of the Ly$\alpha$, we took the peak  position of the red peak. 
	We then evaluated the Ly$\alpha$ velocity shift as the velocity difference between the Ly$\alpha$ redshift and the systemic redshift. The error has been evaluated from the usual propagation formula using the error on $z_{sys}$ and the  error for the Ly$\alpha$ wavelength position evaluated in a similar way. 
	
	In some spectra the wavelength range around the Ly$\alpha$ was  affected by noise,  e.g., due to  residuals from skylines, and therefore it was not possible to derive an accurate   peak position of the Ly$\alpha$ line. We decided to keep these spectra to derive the other spectral parameters, such as  EW(Ly$\alpha$) (see below). In the end we   could measure the Ly$\alpha$ velocity shift   only for 46 galaxies out of 52. We show for reference the Ly$\alpha$ line profile of the galaxy  CDFS122764 in the bottom left panel of Figure \ref{fig:examplefitspectrum}. Here, the Ly$\alpha$ velocity shift is the velocity difference between the peak of the line (dotted line) and the green line that represents the position of the line at systemic redshift.

	\item \emph{The Ly$\alpha$  FWHM}
	
	For the same 46 objects for which we measured the Ly$\alpha$ peak position,  we also evaluated the FWHM of the line,  directly taking the wavelength interval at half of the flux of the  peak of the line.  Again we  prefer a direct measure of the FWHM rather than a Gaussian fit since in many cases the line is clearly asymmetric and the Gaussian is not a good representation of its shape. We estimated the error using the formula in \cite{lenz92}.
	
	\item \emph{The Ly$\alpha$ and CIII] $\lambda$1909$\AA$ EWs}
	
	To estimate the Ly$\alpha$ equivalent width we first normalised the line flux. We used as continuum the mean flux in a spectral range between the Ly$\alpha$ and the SiII at $1260\AA$ that is the first strong absorption line red-ward the Ly$\alpha$ that we can clearly see in the VANDELS spectra\footnote{We did not include possible effects from HI absorption around the Ly$\alpha$ in our estimate of continuum. This might result is a slight underestimate of the EW.}. We then evaluated the EW(Ly$\alpha$) from the integrated normalised flux. To estimate the error on the EW(Ly$\alpha$) we used equation 13 in \cite{ebbets95}.
	We show for reference the Ly$\alpha$ line profile of the galaxy CDFS122764 in the bottom left panel of Figure \ref{fig:examplefitspectrum}, where the shaded region represents the integrated  flux of the line. 
	
	The EW(CIII]) has been evaluated directly from the fit used to derived the systemic redshift, normalizing the line flux with the best fit line of the continuum. As for the EW(Ly$\alpha$) we used the \cite{ebbets95} formula to derive the errors, using as continuum uncertainty the standard deviation in a wavelength range of $100\AA$ observed-frame red-ward the CIII] $\lambda$1909$\AA$ line.

	\item \emph{The LIS redshift ($z_{IS}$) and velocity shift ($IS_{shift}$)}
	
	
	The low-ionization absorption lines (LIS) are among the strongest lines detectable in UV rest-frame spectra of star-forming galaxies. In particular the strongest lines that we can observe in the majority of our spectra are the SiII  $\,\lambda 1260 \AA$, CII  $\,\lambda 1334 \AA$, SiII  $\,\lambda 1526 \AA$, FeII  $\,\lambda 1608 \AA$ and the Al III  $\,\lambda 1679 \AA$. In high redshift star forming galaxies, whose UV continuum is dominated by young stars, these lines are mainly produced by the absorption of the stellar radiation by the ISM and are a fundamental tool to study its properties.
	
	We used the velocity shift of the LIS to probe the  velocity of the ISM. We chose to use only the lines with the highest S/N in our spectra: SiII $\lambda$1260$\AA$, CII $\lambda$1334$\AA$ and SiII $\lambda$1526$\AA$. We evaluated the redshift of these lines using a combined fit of all the three features, even though in some cases we  excluded the lines that did  not have a sufficient S/N (less than 3). We fitted the continuum with a fifth order polynomial function, plus three Gaussian profiles to fit the lines (or less if we could  not fit all the lines) to derive $z_{IS}$, imposing the line centers at $1259.33\,(1+z)$, $1333.95\,(1+z)$ and $1526.13\,(1+z)$, respectively. We again excluded the wavelength regions where other  absorption lines were visible or where strong residuals from sky lines were present.
	
	We then evaluated the LIS  velocity shifts as the velocity difference between the LIS  redshift and the systemic redshift. The error has been evaluated from the error propagation formula using the error on $z_{sys}$ and the  error for the LIS redshift. 
	
	We could estimate the $IS_{shift}$ only for 27 galaxies because the S/N of the LIS for the remaining sources was not sufficient to evaluate the IS velocity. We show  the spectral continuum of the galaxy  CDFS122764 in the top panel of Figure \ref{fig:examplefitspectrum} for reference. The blue line is the VANDELS spectrum and the magenta line is the  part of the spectrum that we used for fitting the LIS and the continuum. For this object we could fit all the three LIS, and we excluded from the fit the strong absorption lines (for instance OIdb $\lambda$1303$\AA$, the SiIV+OIV $\lambda$1397$\AA$ and the CIV $\lambda$1549$\AA$) that would have altered the shape of the continuum.

	\item \emph{The Ly$\alpha$ and UV spatial extensions}
	
	\begin{figure}
		\includegraphics[width=1\linewidth]{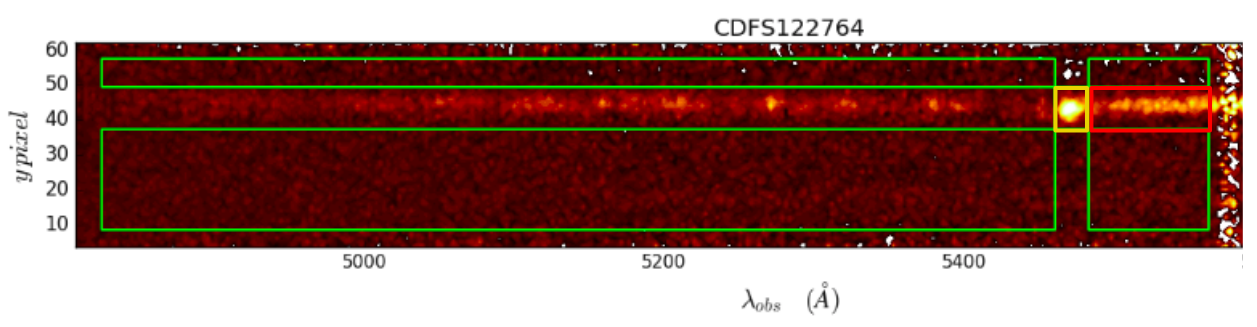}
		\caption{\small{The two dimensional spectrum of the object CDFS122764 that we show to visualize the method that we used to evaluate the Ly$\alpha$ and the UV extents. The green big rectangles show the spatial regions of the two dimensional spectrum that we used to estimate the background, while the yellow (red) square around the Ly$\alpha$ (UV continuum) is the region where we collapsed the spectrum.  }}
		\label{fig:example2dspectrum}
	\end{figure}
	 The Ly$\alpha$ spatial extent was evaluated directly from the two-dimensional spectrum of the sources. 	
	 To estimate the background, we selected four spatial windows on the 2d spectrum, as shown in Figure \ref{fig:example2dspectrum}, which do  not contain the continuum nor the Ly$\alpha$ residuals that can  affect the  regions below and above the Ly$\alpha$ emission in the 2d spectrum. We then separately collapsed the two bottom and upper windows and the region that contains the Ly$\alpha$ and evaluated the mean value for each y-pixel. We then fitted it with a Gaussian profile using a straight line to fit the background and finally evaluated the Ly$\alpha_{ext}$ as the FWHM of the best fit.  To determine the uncertainties in these measurements,  we evaluated not only the error on the Gaussian fit as in  \cite{lenz92}, but also included  the error on the background which we evaluated performing  Monte Carlo simulations by varying the value of each y-pixel within the standard deviation of the background and fitting each time with a Gaussian profile.  The final uncertainties are computed  by combining   the two errors in quadrature, assuming that they are independent. We note that the two different errors are comparable on average.
	
	We evaluated the UV spatial extension in the same way we measured the Ly$\alpha$ extent but using for the continuum the spectral region between the two spatial windows red-ward the Ly$\alpha$ as shown with a red rectangle in Figure \ref{fig:example2dspectrum}\footnote{The typical spatial windows are $\sim25\AA$ observed-frame $\times$ 30 pixels for the Ly$\alpha$ region and $\sim150\AA$ observed-frame $\times$ 10 pixels for the UV continuum. We further checked these regions objects by objects visually inspecting the 2-dimensional spectra.}. 
	
	Our measurements of the Ly$\alpha$ and UV sizes cannot be taken as absolute estimates of these quantities since we did not deconvolve them  by the PSF of the observations. However, this is not a crucial  issue for the present analysis since we are only interested in relative values and given that  the observations were taken under very similar  average conditions over many separate nights, we can assume that the PSF affects all the sources in the same way \citep[see][]{pentericci18}. 
	
\end{itemize}

\subsection{Physical parameters}

\subsubsection{M$^*$, SFR, E(B-V) and  mass-weighted galaxy age}
We performed spectral energy distribution (SED) fitting   to derive the physical parameters of the sources in our sample. For the objects that fall within the CANDELS CDFS and UDS footprints, we used the CANDELS official photometric catalogs \citep{galametz+13,guo+13}, while for the galaxies in the wide-field areas, we used the photometric catalogs described in \cite{mclure18}. Both catalogs span a wavelength range that goes from the U band to the K band observed-frame.
\begin{figure*}
	\centering
	\includegraphics[width=0.68\linewidth]{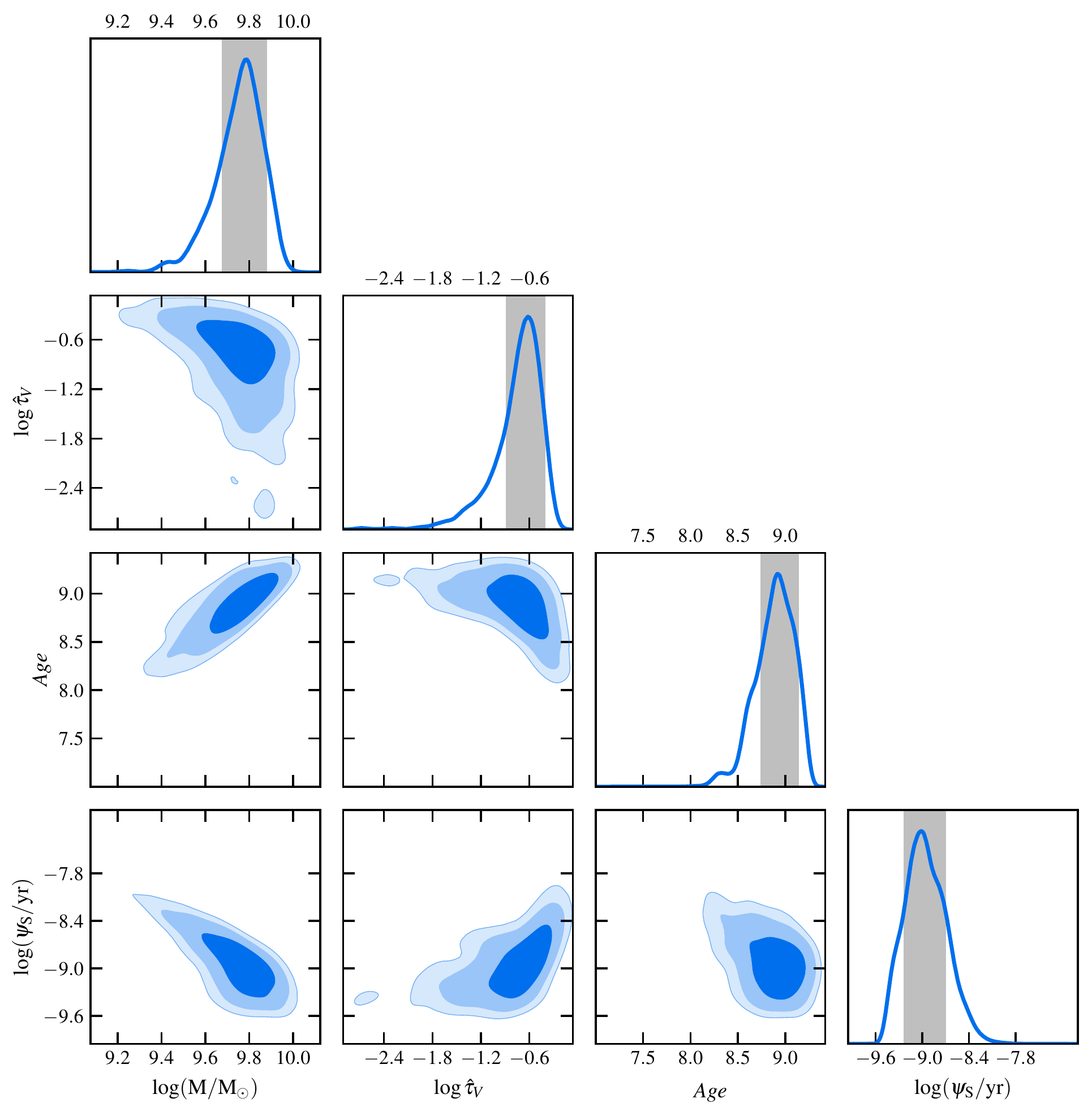}
	\includegraphics[width=0.68\linewidth]{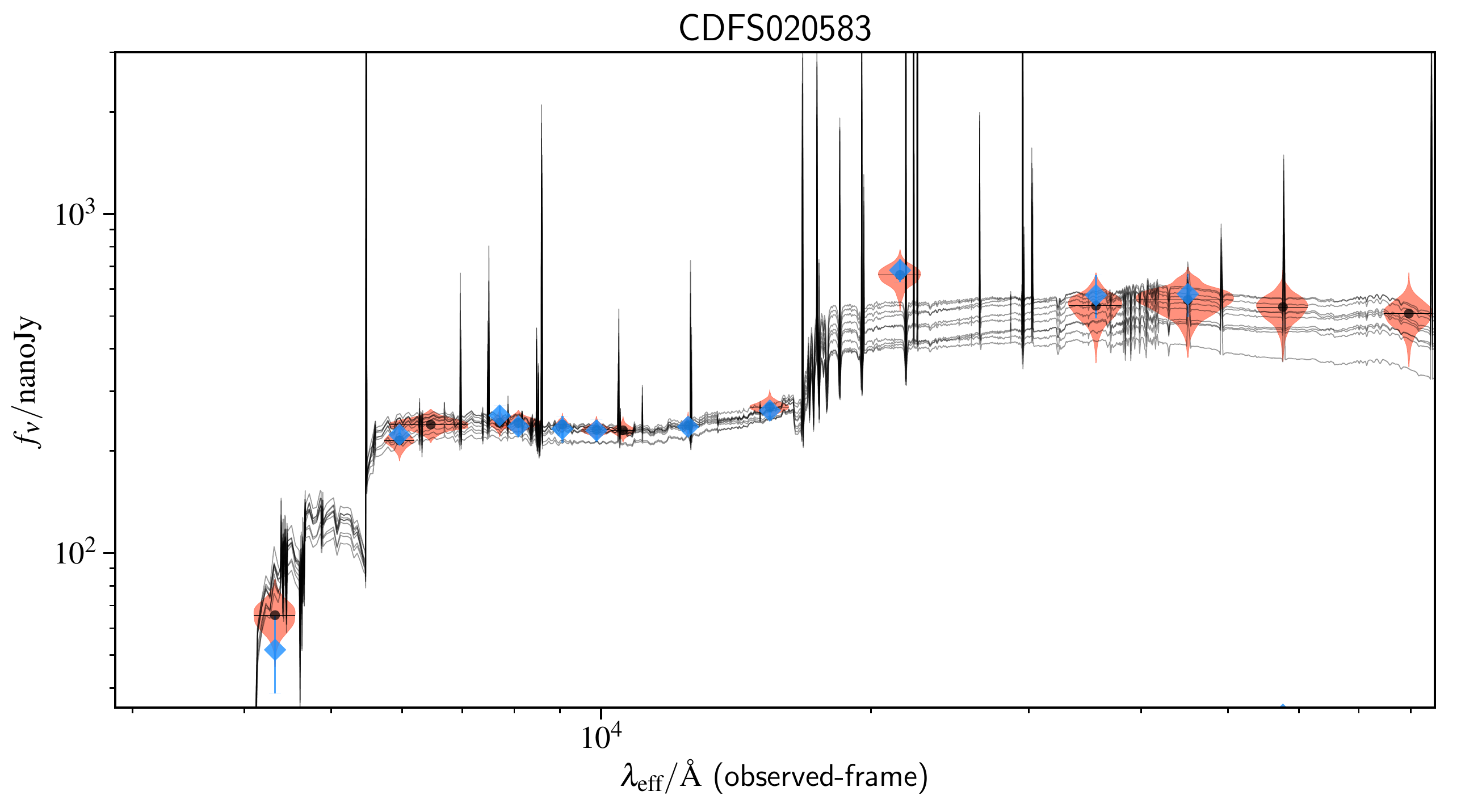}
	\caption{\small{\textit{Top panel}: posterior probability distributions of the physical parameters derived for the object CDFS020583 at $z=3.5$. From left to the right we show: the logarithmic mass in units of solar masses, the dust attenuation coefficient, the age and the logarithmic specific SFR. The different contours indicate the 1$\sigma$, 2$\sigma$ and 3$\sigma$ confidence levels. \textit{Bottom panel}: spectral energy distribution of the same galaxy. The blue points are the photometric observations while the orange shaded regions are the 68\% confidence intervals derived by BEAGLE. The black solid lines are ten among the theoretical SEDs with the highest likelihood values.}}
	\label{fig:SEDbeagle}
\end{figure*}

To perform the SED fitting  we used BEAGLE \citep[for BayEsian Analysis of GaLaxy sEds, ][]{beagle16}, a new-generation tool that allows one to model spectroscopic and photometric galaxy observations with a self-consistent physical model. We used BEAGLE with the most recent  prescriptions of Charlot \& Bruzual (in prep) describing the emission from stars and the models from \cite{gutkin16} to account for emission from photoionized gas. The redshift was fixed at the systemic value that we evaluated from the VANDELS spectra for the objects in the selected sample, while we used the official VANDELS spectroscopic redshift for the galaxies in the parent sample. We used the \cite{chabrier03}  initial mass function  and a \cite{calzetti00} attenuation law, since it was recently shown by \cite{cullen18} that VANDELS  galaxies at $3<z<4$ are consistent with this law. 
We used a constant star formation history (SFH) and fitted the metallicity using  a Gaussian prior centered at $0.14\,Z_{\odot}$\footnote{BEAGLE adopts a solar metallicity of $Z_{\odot}=0.01524$} with a $\sigma$ of $0.07\,Z_{\odot}$. The  value of $0.14\,Z_{\odot}$ corresponds to  the average metallicity of our sample  evaluated applying the method described in \cite{sommariva12} on the stacked spectrum of all the galaxies in the selected sample. Briefly, this method exploits  spectral indices that are present in the UV spectra of star-forming galaxies in the range between $\sim 1300\AA$ and $1600\AA$ rest-frame, whose EWs depend only on the metallicity of the objects. Among the indicators described in \cite{sommariva12}, we used the spectral indices F1370, F1425, F1460 and F1501, because they are not affected by the interstellar absorption lines at our resolution. The value of the standard deviation of the Gaussian prior is the dispersion of the values derived by the four indices. We remark that the aim of this measurement is only to find an average metallicity to lower the number of free parameters in the SED fitting.\\
We derived the stellar masses, star formation rates (SFR), mass-weighted ages and the color excess E(B-V), which  measures  the dust content.  We show  in Figure \ref{fig:SEDbeagle} an example of a SED fit of a galaxy in our sample. In the \textit{top panel} we show the  probability distribution functions of the derived physical parameters and in \textit{bottom panel} we compare the observed photometry with the one predicted by BEAGLE. 
\begin{figure*}
\centering
\includegraphics[width=0.7\linewidth]{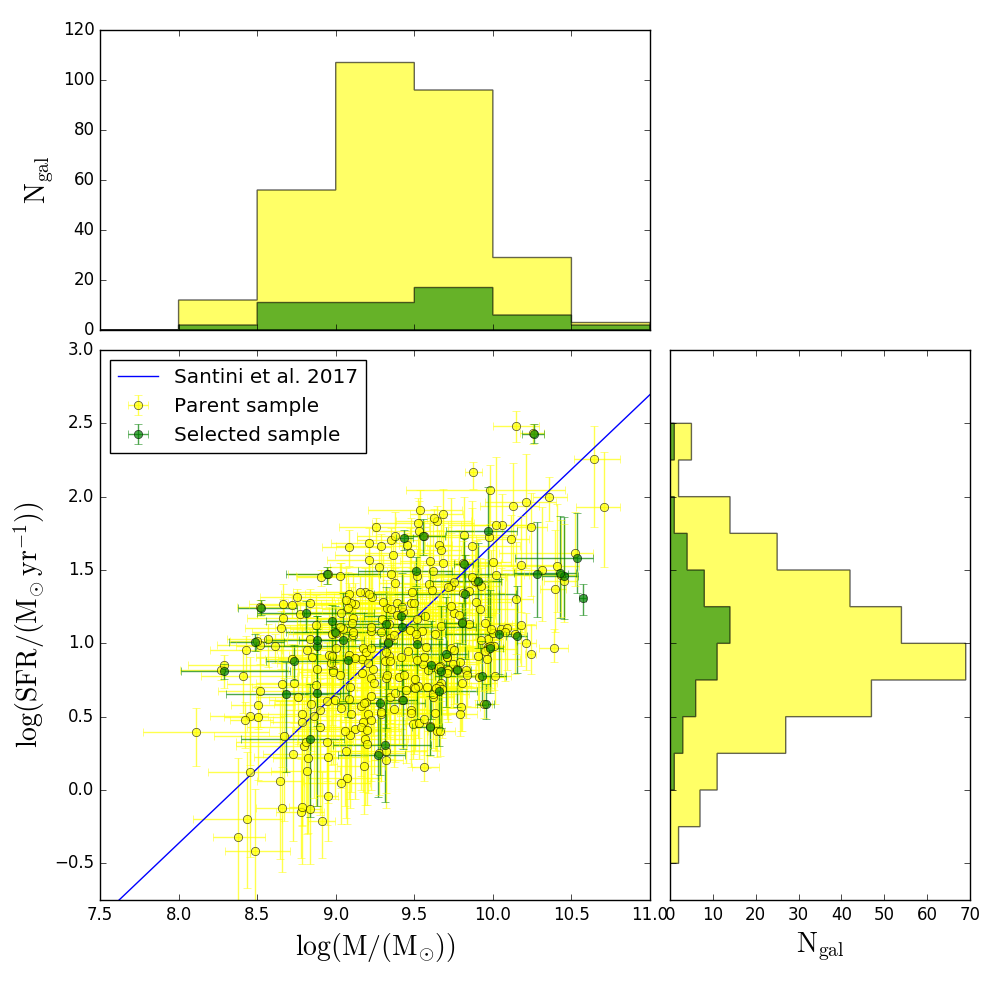}
\caption{\small{SFR as a function of the stellar mass for the objects in our sample (green points) and the parent sample (yellow points). The blue line is one of the latest star formation main sequence evaluated by \cite{santini17} for galaxies in the first four \textit{HST} Frontier Fields in the redshift range $3\lesssim z \lesssim 4$.}}
\label{fig:MvsSFR}
\end{figure*}
We show in Figure \ref{fig:MvsSFR} the SFR as a function of the stellar mass for the sources in our selected sample in green and the parent sample in yellow, along with the star-formation main sequence relation evaluated by \cite{santini17} for star-forming galaxies in the first four \textit{HST} Frontier Fields in the redshift range $3\lesssim z \lesssim 4$ (blue line). We also show in the same Figure the distributions of SFRs and stellar masses for the sources in our sample (green histogram) and in the parent sample (yellow histogram).  Both the samples lie on average on the star formation main sequence, as the majority of high redshift star-forming galaxies. This relation is believed to characterize galaxies that have grown on long timescales as a consequence of smooth cold gas accretion from the IGM \citep[e.g. ][]{dekel09}. We note that there is some scatter around the main sequence in Figure \ref{fig:MvsSFR}, that is however consistent with that of the \cite{santini17} sample. The reason for this scatter could be related to the SFH used in the SED fitting procedure \citep[e.g.][]{cassara16}. To check this, we have  also performed the same SED-fitting  using free metallicity in the range $[0.006\,Z_{\odot};1.74\,Z_{\odot}]$ and two different types of SFH: exponentially declining SFH and delayed exponentially declining SFH. In both cases the scatter around the main sequence is consistent with that obtained with constant SFH. Moreover, also the individual values are consistent. In particular, while the stellar masses are usually not affected by the choice of SFHs, the SFRs can instead vary significantly. In this case however, if we compare the results obtained with constant SFH and exponential and delayed SFHs, they are still consistent within the 1$\sigma$ errors. In a future work we will use BEAGLE to assess the implications of this choice with a larger sample of galaxies in the VANDELS survey.

\subsubsection{$\beta$ slope}

We also determined the  slope $\beta$ of the UV continuum. We adopted the common power-law approximation for the UV spectral range $F_\lambda \propto  \lambda^\beta$ \cite[see e.g.][]{castellano12}, and estimated the  slope $\beta$ by fitting a linear relation  through the observed AB magnitudes: $M_i=-2.5(\beta+2.0)\mathrm{log}\lambda_i + c$, where M$_i$ is the magnitude in the ith filter at effective wavelength $\lambda_i$. The filters we use span the rest-frame wavelength range $\lambda \sim 1250 - 2500\,\AA$, such that the  Ly$\alpha$ emission line is not included in any of the filters.
We divided our sources in two different bins of redshifts: $3\lesssim z\lesssim 3.5$ and  $3.5\lesssim z\lesssim 4.5$, resulting in the observed AB magnitudes I, Z, Y, and J for the first and Z, Y, J and H for the second. As for the SED fitting procedure, we used  the CANDELS official photometric catalogs \citep{galametz+13,guo+13} for the sources in CANDELS, and the photometric catalogs described in \cite{mclure18} for the galaxies in the wide-field areas.

\section{Parent sample versus selected sample}
\label{sec:parentvsselected}
\begin{figure}
	\centering
	\includegraphics[width=\linewidth]{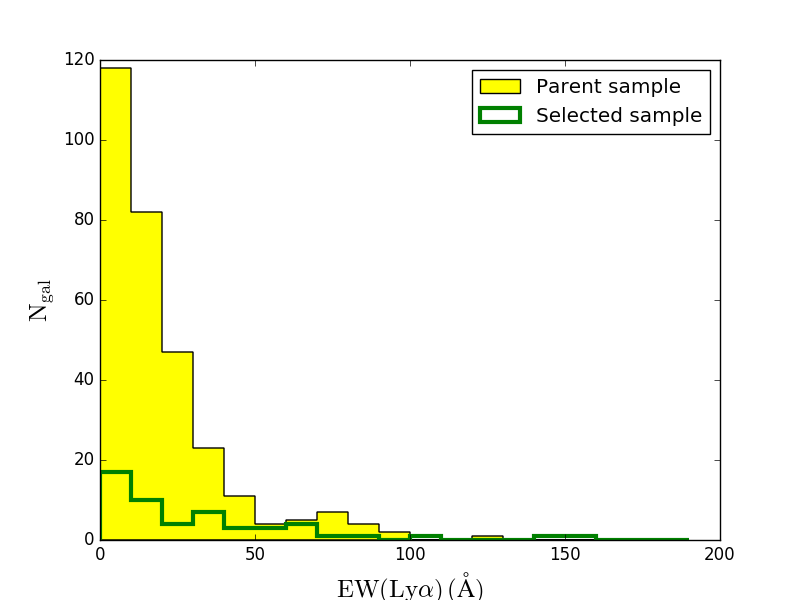}
	\caption{\small{Ly$\alpha$ EW distributions for the sources in the \textit{parent sample} (yellow histogram) and those in the \textit{selected sample} (green empty histogram).}}
	\label{fig:EWhist}
\end{figure}

To determine to what extent our sample (which is mostly based on the presence of the CIII] emission line) is representative of the parent sample of Ly$\alpha$ emitting galaxies, we performed a Kolmogorov-Smirnov test. 
Since we could not check all the $>$ 300 individual objects in the parent sample for the presence of residual of skylines or other extraction problems around the position of the Ly$\alpha$ emission line,  we restricted the parent sample to galaxies where the fundamental parameter, the Ly$\alpha$ EW, was determined with an accuracy of at least $50\%$. 
 We find that, in terms of stellar masses, SFRs, dust content and ages, we cannot reject the hypothesis that  the two samples are drawn from the same distribution.  For the Ly$\alpha$ EW distribution, we find instead a very low p-value ($p=0.002$), meaning that our galaxies are not drawn from the same distribution of the parent sample. The difference  can be clearly appreciated  in Figure \ref{fig:EWhist}, where we show the Ly$\alpha$ EW distribution of the selected sample in green and that of the parent sample in yellow. The median Ly$\alpha$ EWs are $21.4\AA$ and $13.4\AA$ respectively. This difference was  expected  since we selected our sample mostly on the basis of the presence of  the  CIII] $\lambda1909\AA$ which was necessary to evaluate the systemic redshift, and given the apparent correlation between the strength of Ly$\alpha$ and that of CIII] (see below).  However the Ly$\alpha$ EW distribution spans the same range of values as the parent sample, only changing in shape. 
 Very similar considerations can be made for the $\beta$ slopes. The KS test indicates that the two samples are not drawn from the same distributions  ($p=0.06$).  This is  expected given that objects with larger Ly$\alpha$ EW have steeper $\beta$ slopes (see below) and also  the presence of the CIII$\lambda1909\AA$  emission line  might be correlated with a harder ionizing  continuum and therefore a bluer UV spectral slope. Also in this case the two samples still span the same range of $\beta$.

 Because of this, along with the fact that the physical properties of the two samples are drawn from the same statistical distribution, we can reasonably conclude that the results that we find in this analysis for the selected sample will be also valid, at least in qualitative terms, for the general population of galaxies with Ly$\alpha$ emission. This is also corroborated by the fact that, as we show in the next Section,  both samples show the same correlations  between the Ly$\alpha$ EW and the physical properties.

\section{Results}
\label{lyaresults}
\begin{table}
	\centering
	\begin{tabular}{c c c c c}
		\midrule
		\midrule
		parameter1 & parameter2 & $\rho$ & N & significance\\
		\midrule
		$\mathrm{EW(Ly\alpha)}$ & EW(CIII])  & 0.62 & 39 & $>3\sigma$\\
		$\mathrm{EW(Ly\alpha)}$ & age & 0.40 &47 & $>3\sigma$ \\
		$\mathrm{EW(Ly\alpha)}$ & M$^*$  & -0.34  &47 & $\sim 3\sigma$\\
		$\mathrm{EW(Ly\alpha)}$ & E(B-V)  & -0.43 &47 & $>3\sigma$\\
		$\mathrm{EW(Ly\alpha)}$ & $\beta$  & -0.56 & 44 & $> 3\sigma$\\
		$\mathrm{IS_{shift}}$ & Ly$\mathrm{\alpha_{shift}}$ & 0.44 & 23 & $> 2\sigma$\\
		$\mathrm{EW(Ly\alpha)}$ & $\mathrm{UV_{ext}}$  & -0.37 & 38  & $>2\sigma$\\
		$\mathrm{EW(Ly\alpha)}$ & Ly$\mathrm{\alpha_{ext}}$  & -0.39 & 44 & $>2\sigma$\\
		\bottomrule
		\vspace{0.1cm}
	\end{tabular} 
	\caption{Spearman rank correlation coefficients for the different correlations found in this analysis sorted by the significance of the relation and the value of the Spearman rank correlation coefficient. (1) and (2) parameters for which we are testing the (anti)correlation; (3) Spearman rank correlation coefficient, $\rho$; (4) number of sources in the subsets considered; (5) the significance of the relation.}
	\label{tab:spearman}
\end{table}
To study how the kinematics and appearance of Ly$\alpha$ emission depend on the galaxies physical properties and the kinematics of the ISM, we investigated the relations between the galaxy spectral and physical properties evaluated in the previous section. 
In particular, to give a non parametric statistical measure of the correlations between the different parameters, we evaluated the Spearman rank correlation coefficients ($\rho$). 
We show in Table \ref{tab:spearman}  the Spearman rank correlation coefficients for the variables that we find (anti)correlated, and the significance of these relationships. We trust the correlation only if the significance is higher than $2\sigma$.

\subsection{Correlations between Ly$\alpha$ EW and physical properties} 
\label{sub:corrphys}

\begin{figure}
	\centering
	\includegraphics[width=1\linewidth]{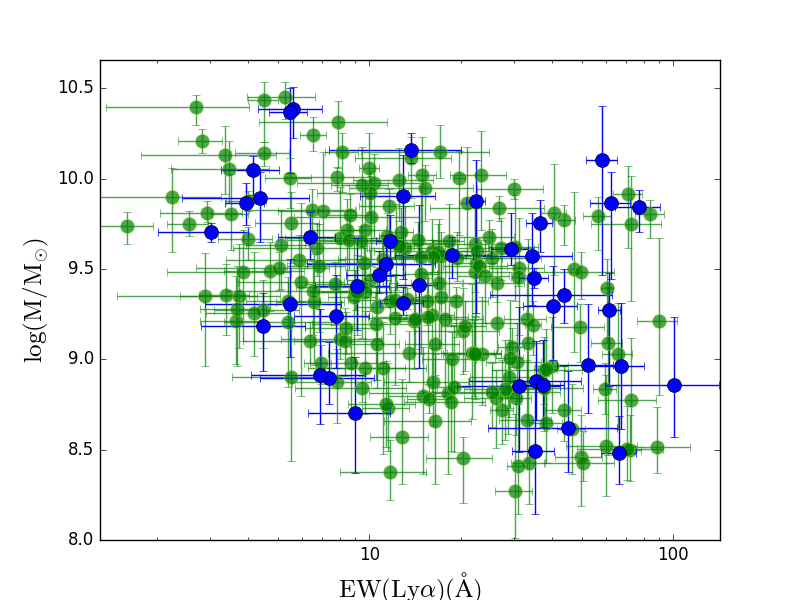}\\
	\includegraphics[width=1\linewidth]{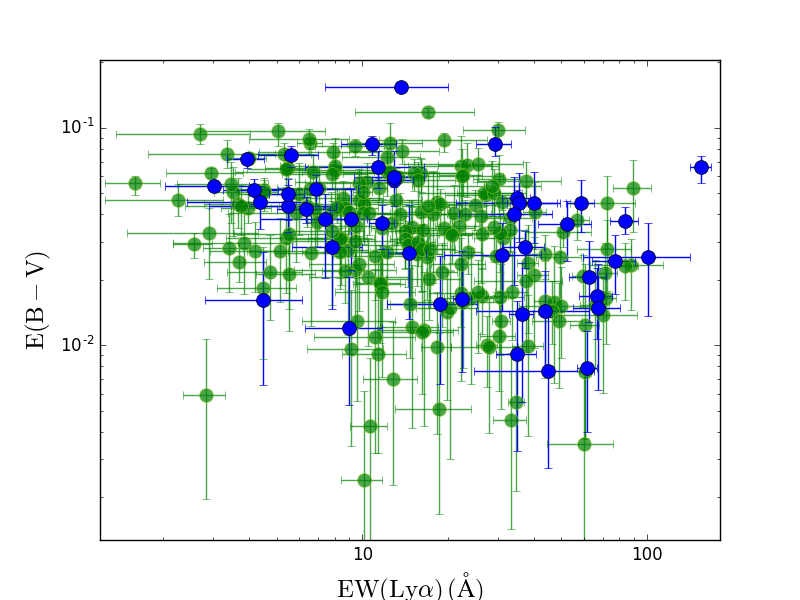}\\
	\caption{\small{Correlations between the Ly$\alpha$ EW and the stellar mass (top panel) and between the Ly$\alpha$ EW and the color excess E(B-V) (bottom panel) for the selected sample (blue) and the parent sample (green). }}
	\label{fig:correlationsEW_phot}
\end{figure}

In this subsection we discuss the main correlations that we find between the Ly$\alpha$ EW and the galaxies' physical properties. We first discuss the results obtained for the selected sample, then we compare them with the corresponding relations for the parent sample.

\begin{itemize}
	
		\item{\textbf{EW(Ly$\alpha$) vs M* }}
		
		We show in Figure \ref{fig:correlationsEW_phot} (blue points) the relation between EW(Ly$\alpha$) and stellar mass M$^*$: the two parameters are moderately anti-correlated  with a Spearman rank correlation coefficient of $-0.34$. 		
		While the lack of objects with large masses and large Ly$\alpha$EW is surely real, we might be losing objects with small masses and small EWs (lower left part of the figure)  because of the flux limit of the survey. We checked that this not the case, given the sensitivity of our observations to emission lines and the median R band magnitude of our targets.
		
		 The existence of this relation is quite controversial since it has been previously observed by some works \citep[e.g.][]{finkelstein07,jones12,hathi16} but not by others \citep[e.g.][]{kornei10}. In particular \cite{pentericci07,pentericci10} found that galaxies without Ly$\alpha$ in emission tend to be more massive and dustier than the rest of the LBGs sample and  found a significant lack of massive galaxies with high EW(Ly$\alpha$), which could be explained if the most massive galaxies were either dustier and/or if they contained more neutral gas than less massive objects. However they did not find a precise one-to-one correlation.   Finally, \cite{du18} found a clear relation for stacks of spectroscopically confirmed LBGs at $z\sim 4$ but they showed that this becomes weaker as redshift decreases with a almost flat trend at $z\sim 3$. 
		
		\item{\textbf{EW(Ly$\alpha$) vs E(B-V)} }
		
		As shown in the bottom panel of  Figure \ref{fig:correlationsEW_phot},  we also observe a moderate anti-correlation between dust extinction,  indicated by the color excess E(B-V),  and EW(Ly$\alpha$) with a Spearman rank correlation coefficient of $-0.43$, in agreement with several previous studies \citep{shapley03,kornei10,pentericci10,jones12,erb16, hathi16}. This is also consistent with theoretical expectations:  Ly$\alpha$ emission is easily quenched by dust, since Ly$\alpha$ photons can be absorbed by dust grains, that then re-emit them as thermal emission.

		\item{\textbf{EW(Ly$\alpha$) vs galaxy age}}
		
		We observe a  correlation between Ly$\alpha$ EW and the mass-weighted galaxy age, in the sense that galaxies with  a higher Ly$\alpha$ EW, are  older than galaxies with lower Ly$\alpha$ EW. The Spearman rank correlation coefficient is $0.40$. While the relation between Ly$\alpha$ and E(B-V) is now quite well established,  the relation  between  Ly$\alpha$ and galaxy age is still highly debated.  Some studies claim that
		Ly$\alpha$ emitting galaxies are  mostly young  galaxies in the very early stages of formation \citep[e.g.][]{pentericci07,du18}, while others  claim that Ly$\alpha$ emission is typically found in older galaxies \citep[e.g.][]{shapley03,kornei10}. Finally, other studies do not find apparent relation between Ly$\alpha$ EW and age \citep[e.g.][]{pentericci09,pentericci10}.	\\
		According to our observations, the brightest Ly$\alpha$ emitters are also the oldest. This could be explained by a scenario in which the oldest galaxies  have  experienced in the past strong burst of star formation   that caused massive stellar winds with the consequent expulsion of neutral gas and dust, resulting in a reduced covering fraction and therefore a smaller attenuation of the Ly$\alpha$ photons, while the younger galaxies still have more dust and therefore smaller Ly$\alpha$ EWs.\\			
		We remark that the ages are one of the most uncertain parameters in SED fitting also because they are somehow dependent on the SFH adopted \citep[see e.g.][]{carnall18,leja18}:  we tested the exponentially declining SFH and the delayed exponentially declining SFH instead of a constant one, but we do not find significant differences in the observed correlation.
		
		\item{\textbf{EW(Ly$\alpha$) vs SFR}}
		
	We do not find any correlation between the Ly$\alpha$ EW and the SFR, measuring a Spearman rank correlation coefficient of 0.14, with very high confidence. This might look in contrast with the results from several previous studies \citep[e.g.][]{kornei10,hathi16} that found instead that these two quantities are anti-correlated, with the brightest LAEs having lower SFRs. However, all  these studies analysed  galaxies with Ly$\alpha$ both in absorption and in emission, while we have only selected galaxies with Ly$\alpha$ emission. Indeed, if we restrict previous results to  only those sources with Ly$\alpha$ in emission, no clear  correlations are observed. This suggests that galaxies with Ly$\alpha$ emission have on average smaller SFRs if compared to the non emitting galaxies, but that there is no particular trend with the actual strength of Ly$\alpha$. Note also that while the sources presented in \cite{hathi16} are in the same mass range as those analysed in this paper, the sources in  \cite{kornei10} have much larger average stellar masses ($\mathrm{log(M/M\odot)\sim 9.9}$).

		\item{\textbf{EW(Ly$\alpha$) vs $\beta$ slope}}
		
		\begin{figure}
			\centering
			\includegraphics[width=1\linewidth]{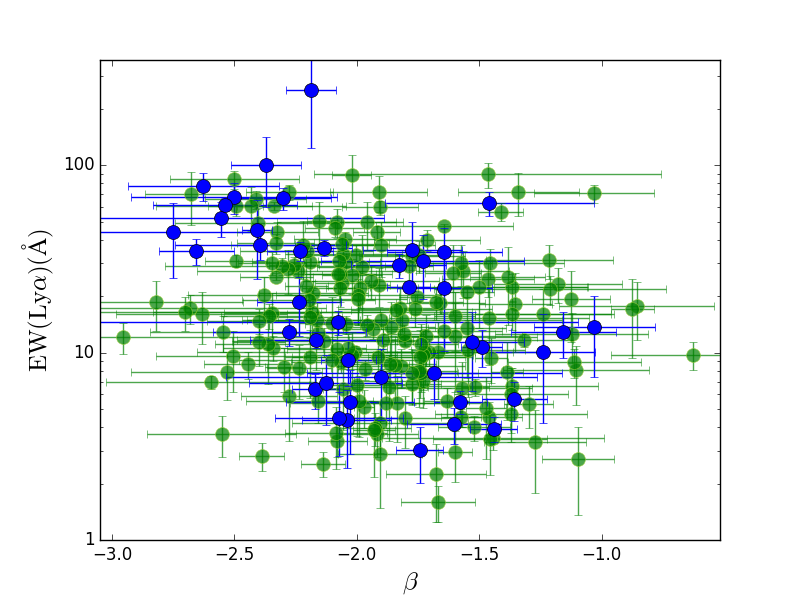}
			\caption{\small{ Correlation between the $\beta$ slope and the  Ly$\alpha$ EW. }}
			\label{fig:correlationsEWbeta}
		\end{figure}
		We reproduce  the well known  anti-correlation between the UV spectral slope $\beta$ and the Ly$\alpha$ EW, with a Spearman rank correlation coefficient of $-0.44$ and high significance (>3$\sigma$). This relation is shown in Figure $\ref{fig:correlationsEWbeta}$ for the sources in the selected sample in blue and for those in  the parent sample in green.
	
		The UV $\beta$ slope depends on several parameters including dust, age and metallicity, although it is thought  to be primarily driven by the dust content \citep[e.g.][]{reddy10,hathi16}. 
		The anti-correlation between  Ly$\alpha$ EW and the $\beta$ slope, is therefore a direct consequence of the fact that less dusty galaxies tend to show brighter Ly$\alpha$ emission as also probed by the anti-correlations that we observe between  the Ly$\alpha$ EW with  E(B-V). 		
	\end{itemize}
	
We finally tested if the relations between EW(Ly$\alpha$) and the physical properties are also valid for the parent sample, that contains galaxies with  similar physical properties but with different distributions of Ly$\alpha$ EWs and $\beta$ slopes. The results of this analysis are also shown in Figure \ref{fig:correlationsEW_phot}. The correlation  between EW(Ly$\alpha$) and stellar mass  and E(B-V) are comparable  to those found for the selected sample, and similarly there is no relation 
between SFR and EW(Ly$\alpha$).
We however do not observe any correlation between the Ly$\alpha$ EW and the galaxy mass-weighted age for the galaxies in the parent sample,   that was  instead present for the sources in the selected sample. We remark  however that, as already mentioned above, the galaxy age is the most uncertain parameter from SED fitting  and that the Spearman rank correlation coefficients and their significance, are evaluated without taking  into account the uncertainty in the individual parameters.
Finally, a correlation between the Ly$alpha$ EW and $\beta$ slope is observed also for the parent sample, even if it is weaker than that observed for the selected sample ($\rho=0.2$ with $>2\sigma$ significance).
	 
	\subsection{Correlations between Ly$\alpha$ EW and spectral properties} 
	
	We discuss in this subsection the relations obtained between the Ly$\alpha$ EW and the other galaxy spectral properties.
	
	\begin{itemize}
			
		\begin{figure}
			\centering
			\includegraphics[width=\linewidth]{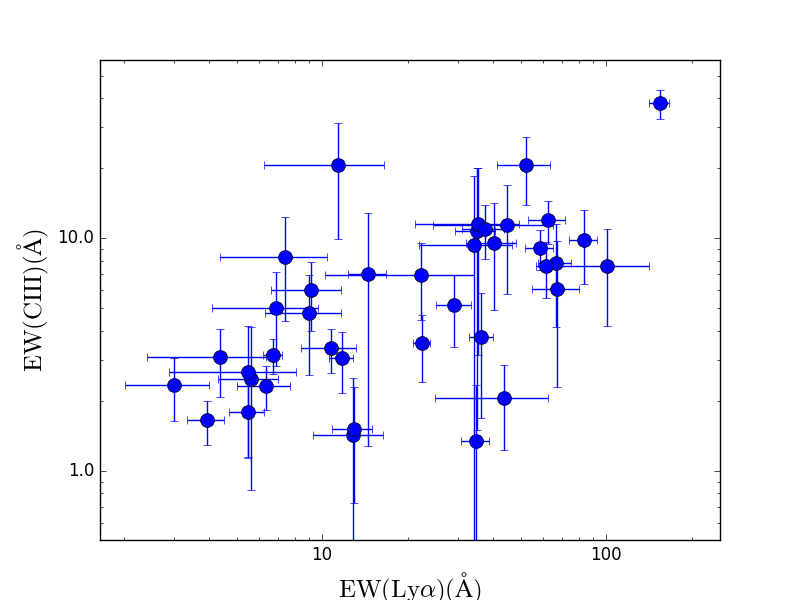} \\
			\includegraphics[width=\linewidth]{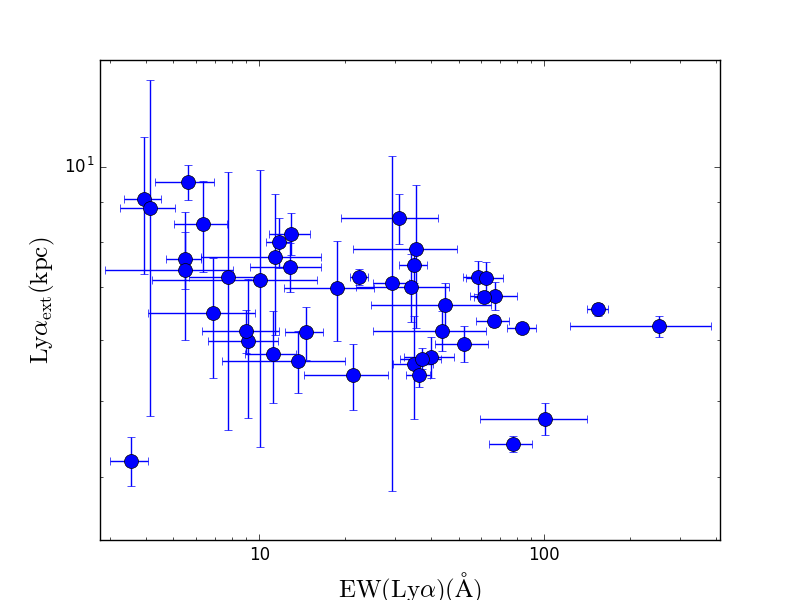}\\
			
			\caption{\small{ Correlations between the Ly$\alpha$ EW and the CIII] EW (top panel) and between the Ly$\alpha$ EW and  the Ly$\alpha$ spatial extent (bottom panel). }}
			\label{fig:correlationsEW}
		\end{figure}
	\item{\textbf{ EW(Ly$\alpha$) vs  EW(CIII])}} 
	
	\label{sub:CIIILyA}
	We observe a strong correlation between EW(Ly$\alpha$) and EW(CIII]), with  Spearman rank correlation coefficient of 0.62, that we show in the top panel of Figure \ref{fig:correlationsEW}. This relation has been previously investigated by other studies, using both stacks and individual spectra with somewhat contradictory results.  \cite{shapley03} found a positive  trend between the two quantities on spectral stacks of a sample of LBGs at $z\sim 3$.  More recently \cite{stark14}  found a tighter relation between EW(Ly$\alpha$) and EW(CIII]) in individual galaxies in a sample of 17 strongly lensed low-mass galaxies at $z\sim 2$. 
	In contrast to these results,  other studies just show tentative correlations with a very  large  scatter. For instance, \cite{rigby15} studied a sample of galaxies at $1.6 < z < 7$, and found an apparent correlation, though with appreciable scatter especially for the weakest Ly$\alpha$ and CIII] emitters ($EW(Ly\alpha)\lesssim 50 \AA$ and $EW(C III])\lesssim 5 \AA$). A very similar large scatter was found by \cite{lefevre17} for individual star-forming galaxies at $2 < z < 3.8$ with CIII] emission in VUDS, with  a clear trend that is only observed  using stacks as in \cite{shapley03}. Finally, we mention that \cite{du18}, using stacks of $z\sim 2$ star-forming galaxies  observed a nearly null correlation  between EW(CIII]) and EW(Ly$\alpha$), with the exception of the strongest Ly$\alpha$ bin where they observe a higher EW(CIII]) with respect to the other sub-samples.
	All these works suggest that, on average, the presence of Ly$\alpha$ (CIII]) emission increases the probability of detecting also CIII] (Ly$\alpha$) emission, but that there is not a clear one to one relation. There are indeed numerous examples of galaxies with CIII] emission that show only a weak Ly$\alpha$ emission or no emission at all, and several LAEs that do not show any sign of CIII] $\lambda$1909$\AA$ emission line \citep[e.g.][]{lefevre17}. This behavior is not surprising. The production of both Ly$\alpha$ and CIII] photons strongly depends indeed on the on-going star formation but there are significant differences in the production and radiative transfer processes of the two lines \citep[see e.g.][for the CIII production]{nakajima18}.
	In our data we find a rather strong correlation  between the Ly$\alpha$ EW and CIII] EW  in  individual galaxies. \footnote{The correlation is unchanged if we exclude the point in the upper right part of Figure \ref{fig:correlationsEW}, that might appear to drive it.}	
	We note that our sample contains galaxies with higher stellar masses compared to e.g. \cite{stark14} who probe a mass range of $9\times 10^6 - 1.3\times 10^9 M_{\odot}$.
	Clearly the tight relation that we observe is somehow biased since we pre-selected the sample to have both Ly$\alpha$ and CIII] emission. We are therefore missing the objects with CIII] (Ly$\alpha$) emission and not Ly$\alpha$ (CIII]). Indeed a similar result was also found by \cite{amorin17} for a sample of 10 low-mass galaxies at $z\sim 3$, selected by their strong Ly$\alpha$ emission ($EW(Ly\alpha) > 45 \AA$) in VUDS. In a future work we will exploit the whole VANDELS dataset to evaluate the unbiased relation between Ly$\alpha$ and CIII]. 
	
	\item{\textbf{EW(Ly$\alpha$) vs  Ly$\alpha_{ext}$ and UV$_{ext}$}} 
	
	We observe a mild anti-correlation between EW(Ly$\alpha$) and the UV and Ly$\alpha$ spatial extents, in the sense that galaxies with higher EW appear to be more compact in size both in the UV continuum and in Ly$\alpha$.  The  Spearman rank test gives correlation coefficients of -0.37 and -0.39, respectively. The relation between Ly$\alpha$ EW and  UV size has been previously observed in Lyman Break galaxies  \citep[e.g.][]{jones12}. The relation with Ly$\alpha_{ext}$ has been previously observed by \cite{guaita17}, using stacks of VUDS galaxies at $z\sim 3$, although their definition of Ly$\alpha$ extension  was slightly different than ours since they  subtracted in quadrature to the Ly$\alpha_{ext}$, the UV continuum to implicitly consider the deconvolution with the PSF of the observations.  We now find  a similar  correlation also using individual objects.   
	

\end{itemize}

We tested these correlations also on a small sub-sample of 17 galaxies, for which we could measure all the quantities listed in Section \ref{lyamethod}. We find for this sample the same correlations observed before and with the same significance.

\subsection{Correlations with IS shift} 

\begin{figure*}
	\centering
	\includegraphics[width=0.49\linewidth]{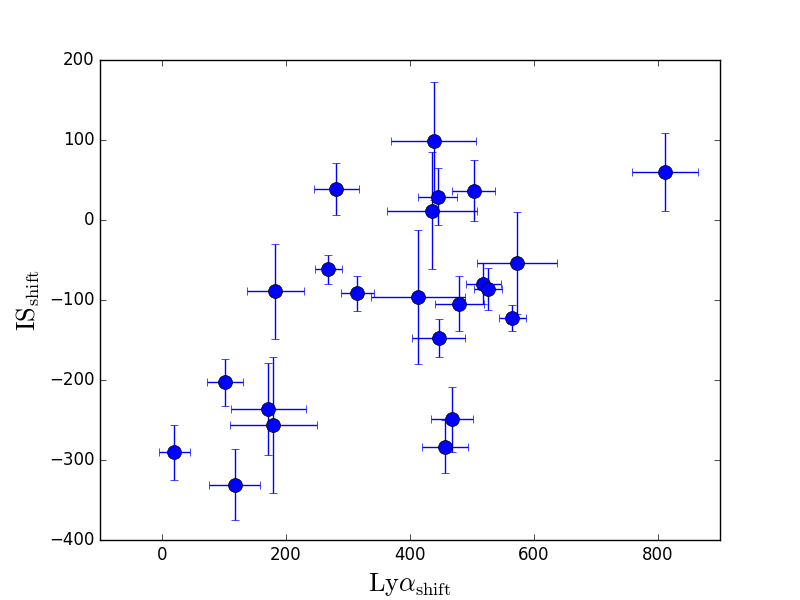}
	\includegraphics[width=0.49\linewidth]{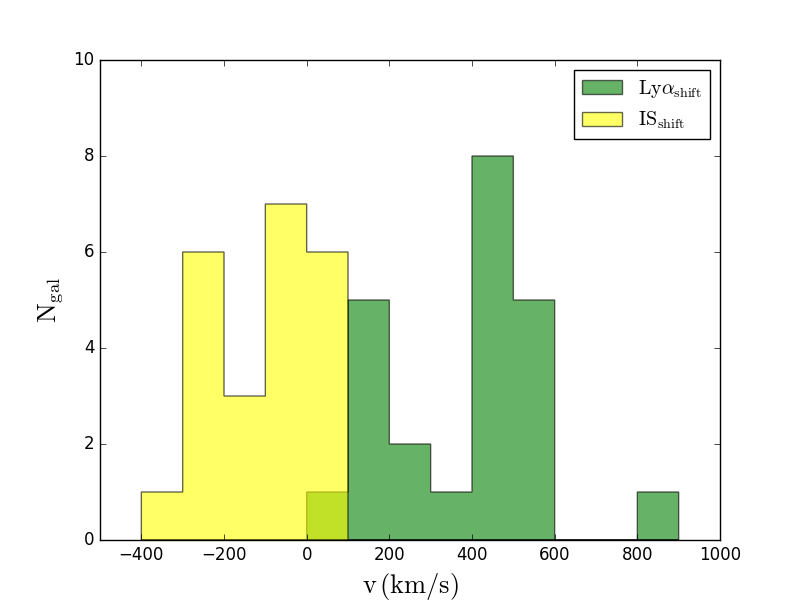}
	\caption{\small{ \textit{Left panel:}  anti-correlation between the velocity of the main peak of the Ly$\alpha$ profile with respect to the systemic redshift and the IS velocity shift.  \textit{Right panel:} Histogram of the measured (centroid) velocities of interstellar absorption lines in yellow and Ly$\alpha$ emission in green with respect to the galaxy systemic redshift derived from the centroid of either the CIII]$\lambda 1909\AA$ or the HeII$\lambda 1640\AA$ emission lines. The yellow sample contains only the 27 objects for which we could give a precise measure of the IS velocity shifts, and the green sample contains the 46 spectra where we could estimate the Ly$\alpha$ peak shift. The median values of the velocity offsets are $IS_{shift}=-89\, km/s$ and $Ly\alpha_{shift}=400\,km/s$, respectively.}}
	\label{fig:correlationsIS}
\end{figure*}

We tested if the velocity of the IS medium is related to other galaxy parameters.  The only property that we find related to the $\mathrm{IS_{shift}}$, is the offset of the Ly$\alpha$ line with respect to the systemic redshift, that we show in Figure \ref{fig:correlationsIS}, along with the distributions  of the two parameters, in yellow and green, respectively.

The Spearman rank correlation coefficient for this relation is 0.44, with a significance that is $> 2\sigma$. This relation has been previously observed by  \cite{guaita17} using stacks of $z\sim 3$ star forming galaxies.  We discuss this trend and its interpretation  in the next Section.

We  do not observe any correlation between IS shift and other physical parameters. In particular while we expected some trend with the SFR,  the Spearman rank correlation coefficient is only -0.15, with a significance that is  $<<2\sigma$. This trend is well established  in low redshift samples \citep[e.g.][]{weiner09,heckman15} but not at high redshift. In principle, a relation  would be expected also at high redshift since outflows are believed to be driven by  pressure created by supernovae, stellar winds, and the radiation field caused by star formation.  However our  SFRs are derived from SED fitting and have  uncertainties: to properly test this relation, SFR derived from H$\alpha$ line flux would be needed.
Besides SFR, we also specifically tested the correlation between  outflow velocity and  SFR surface density, as defined in \cite{heckman11},
and between outflow velocity and specific star formation rate  defined as $\mathrm{SFR/M^*}$ but again we find no significant trend in our sample.  


\section{Discussion:  the origin of the relation between ISM and Ly$\alpha$ velocity shifts}
\label{lyadiscussion}

The correlation observed in  Figure \ref{fig:correlationsIS} is the most interesting result that we find in this analysis since  it relates the Ly$\alpha$ line kinematics to that of the ISM.
\begin{figure}
	\centering
	\includegraphics[width=\linewidth]{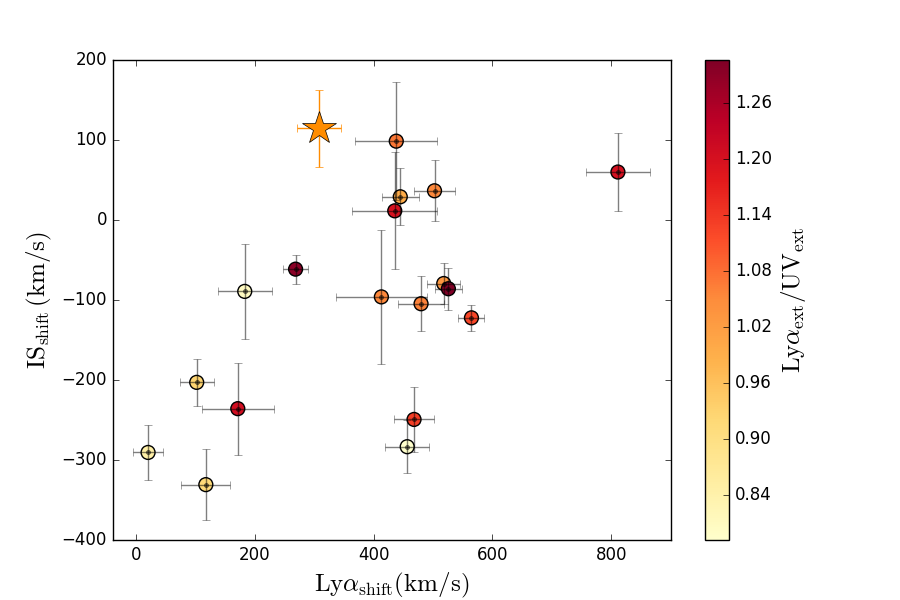}
	\caption{\small{Ly$\alpha$ velocity offset vs IS velocity offset with respect to the systemic redshift. The plot is color coded for the Ly$\alpha$ spatial extension with respect to the UV continuum, as shown from the color bar in the right side of the Figure. The star on the plot is the stack of the sources with no individual $\mathrm{IS_{shift}}$ measure. Its colour is given by the median of the measured $\mathrm{Ly\alpha_{ext}/UV_{ext}}$ of the IS undetected sample.}}
	\label{fig:correlationsLyAIS}
\end{figure}
According to theoretical predictions, the Ly$\alpha$ kinematics  depends on the velocity of the ISM, but also on the HI column density,  that gives a measure of the amount of neutral gas along the line of sight. In this Section we interpret the relation found between Ly$\alpha$ velocity and ISM outflow in the context of the theoretical model of \cite{verhamme06}, that accurately predicts the Ly$\alpha$ shift, Ly$\alpha$ EW and spatial extension as functions of the ISM kinematics and HI column density. In this framework galaxies are modeled as a central region of young stars surrounded by an expanding, spherical, homogeneous, and iso-thermal shell of neutral hydrogen. 
In this scenario, small Ly$\alpha$ shifts and large (but less than $\sim 300\, km/s$) outflow velocities are predicted only for  small HI column densities ($\sim10^{19}\, cm^{-2}$), while large Ly$\alpha$ shifts but with small or null  outflow velocities can be observed only in case of large HI column densities  ($\sim10^{20}- 10^{21} cm^{-2}$) because the Ly$\alpha$ photons must have undergone many scatters before escaping the galaxy, resulting in higher Ly$\alpha$ velocity offsets with respect to the systemic redshift. According to this, we expect the objects in the lower left part of Figure $\ref{fig:correlationsIS}$ to be characterized by smaller HI column densities, while the objects in the upper right part of the plot, by larger HI column densities. 

\begin{figure*}
	\centering
	\includegraphics[width=0.49\linewidth]{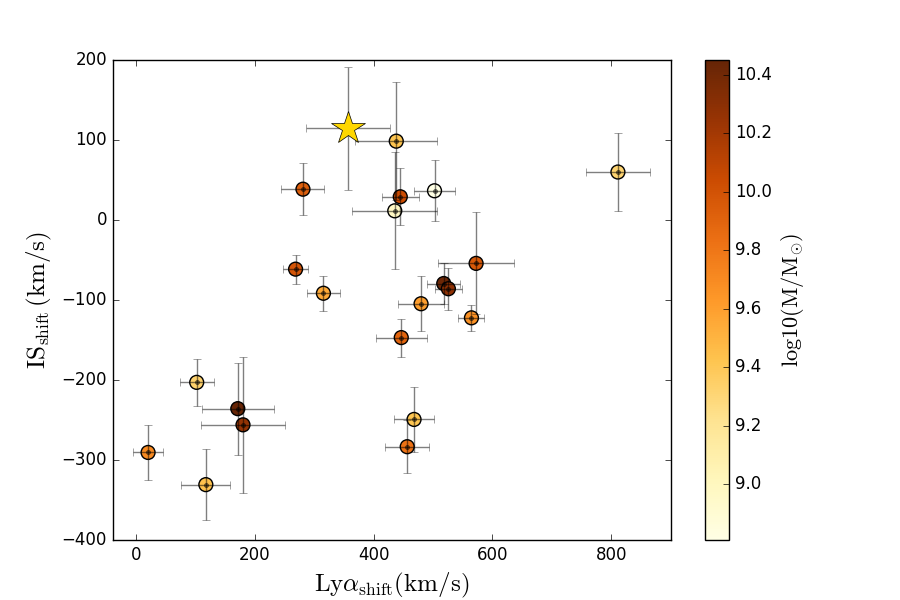}
	\includegraphics[width=0.49\linewidth]{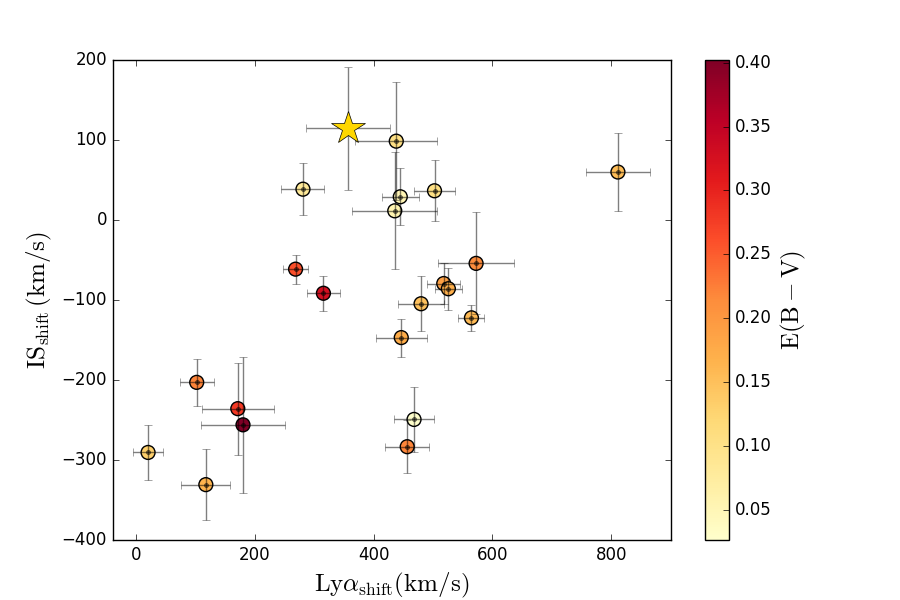}
	\includegraphics[width=0.49\linewidth]{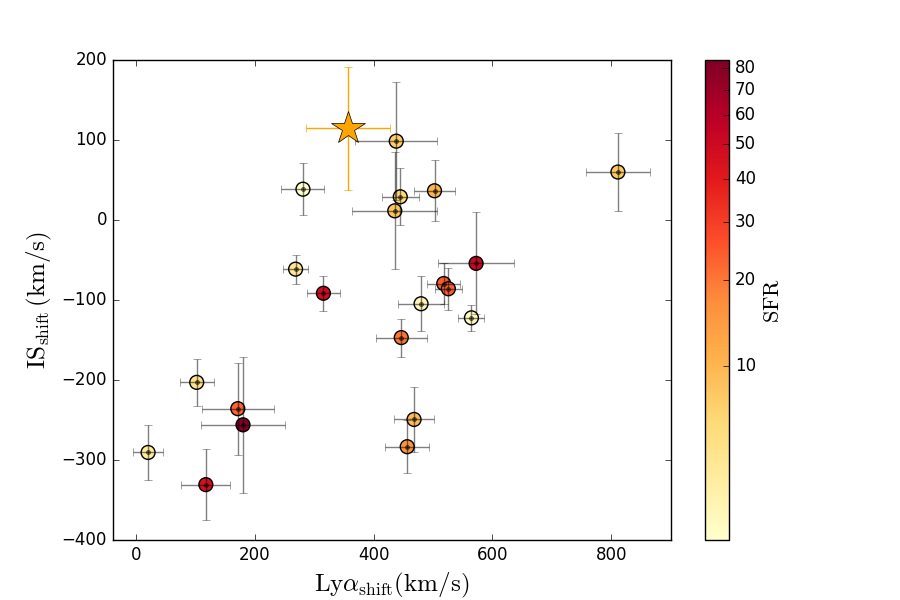}
	\includegraphics[width=0.49\linewidth]{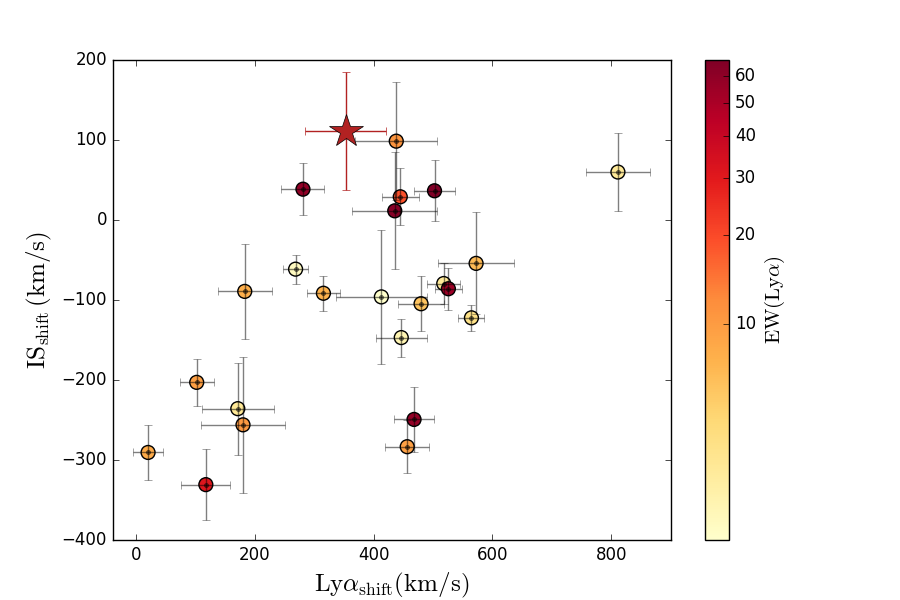}
	\caption{\small{$\mathrm{IS_{shift}}$ vs $\mathrm{Ly\alpha_{shift}}$ color coded for different galaxy properties. From left to the right and top to bottom: stellar mass,   E(B-V) color excess, SFR and EW(Ly$\alpha$). The star on each plot is the stack of the sources with no individual $\mathrm{IS_{shift}}$ measure, with the colour given by the median values of the measured quantity that is indicated in the color bar, for the galaxies in the IS undetected sample.}}
	\label{fig:correlationsLyAIScc}
\end{figure*}

This scenario  is also  consistent with  the distribution of  the Ly$\alpha$ spatial extension with respect to the UV continuum  of our sources.  We show in Figure \ref{fig:correlationsLyAIS} the same plot as in Figure \ref{fig:correlationsIS}, but color coded for the Ly$\alpha$ spatial extent with respect to the UV extension ($\mathrm{Ly\alpha_{ext}/UV_{ext}}$). The objects that appear to have   lower HI column densities (from the Ly$\alpha$ vs IS kinematics) also show on average more compact Ly$\alpha$ spatial profile, while the objects  that should have larger HI column density  (from the Ly$\alpha$  kinematics) also have 
large  Ly$\alpha$  extension compared to the UV extensions. These observations are perfectly in agreement with the shell model predictions, since broader Ly$\alpha$ spatial profile with respect to the continuum are expected if the Ly$\alpha$ photons  undergo many scatters before escaping the galaxy (caused by larger HI column densities), while more compact spatial profile are predicted for galaxies with lower HI column densities, since the Ly$\alpha$ photons can escape galaxies in a more direct way (Verhamme et al. in preparation). 
Assuming that this model is actually working,  and that the trend  observed in Figure \ref{fig:correlationsIS}, is due to a different HI column density,  we then try to  assess if there is any apparent  relation between N(HI) and other galaxy  physical  properties.
In particular, we show in Figure \ref{fig:correlationsLyAIScc}, the same relation between $\mathrm{IS_{shift}}$ and $\mathrm{Ly\alpha_{shift}}$ color coded for the stellar mass, the attenuation of the dust, indicated by the color excess E(B-V), the SFR and  the Ly$\alpha$ EW.  None of these  physical parameters  appears to correlate with the HI column density, i.e. we do not find any significant change  of physical properties in the galaxies as we move from the lower-left part of the figure (small HI column densities) to the upper-right part (high HI column densities). A possible limitation of this analysis is  that we are probing only a relatively  small range for these quantities.  For  example our galaxies span only a factor of 10  in mass: it  is possible that if a weak correlation between N(HI) and mass  exist, it would emerge only by  probing much larger ranges of values.  

We also do not observe any trend between N(HI) and Ly$\alpha$ EW. A tentative anti-correlation between these quantities has been observed  in local galaxies \citep[e.g.][]{pardy14,yang+17}. From a theoretical point of view, this would be expected only if the intrinsic EW  and/or the gas to dust ratio were approximately constant. In the first case, with the increase of N(HI), the path length of Ly$\alpha$ photons would increase and so their chance of being absorbed by dust, while in the second case an increase in N(HI) would also lead to an increase in the dust optical depth, suppressing the Ly$\alpha$ EW. We do not observe this relation. This is however not in contrast with our results since we do not have reasons to believe that the conditions to observe this relation are met in our sample of galaxies. From Figure \ref{fig:correlationsLyAIScc} we do not observe indeed any relation between dust attenuation and the HI column density, suggesting a non constant gas to dust ratio. Moreover, we do not observe any correlation between the Ly$\alpha$ peak shift and EW(Ly$\alpha$), 
relation that would be expected only if the Ly$\alpha$ EW was correlated with  N(HI), since larger HI column densities cause more scattering of the Ly$\alpha$ photons, shifting the line peak to redder wavelengths.

What we  actually observe from Figure \ref{fig:correlationsLyAIScc} is that galaxies with brighter Ly$\alpha$ emission show on average  no outflows and Ly$\alpha$ peak shifts around 400 km/s. It is not yet clear if there is a real physical motivation under these observations and this is still under investigation.

\section{Summary and conclusions}
\label{summaryLyACIII]}
In this work we have investigated the correlations between the physical properties, the Ly$\alpha$ properties and the kinematics of the inter-stellar medium in a sample of 52 star-forming galaxies selected to have Ly$\alpha$ emission and systemic redshift from the VANDELS survey. We find that the Ly$\alpha$ EW and the CIII] $\lambda$1909$\AA$ EW are strongly related, in agreement with  previous observational studies at similar redshifts \citep[e.g.][]{stark14}, suggesting an intrinsic connection between the production of the two lines. We note however that for this correlation the sample is biased, since we only selected galaxies with both Ly$\alpha$ and CIII]  $\lambda$1909$\AA$. This relation could therefore change if the entire population of galaxies with CIII]$\lambda$1909$\AA$  emission would be considered. We also observe a good anti-correlation, although with some scatter, between the Ly$\alpha$ EW and the UV and Ly$\alpha$ spatial extents, with compact sources being the brightest Ly$\alpha$ emitters.  The Ly$\alpha$ line appears also brighter in lower mass and less dusty galaxies, while no correlation appear to exist with the SFR. We finally observe a very interesting correlation between the   Ly$\alpha$ velocity shift and IS shift in the sense that  galaxies with large ISM outflow velocities also show small Ly$\alpha$ velocity shifts. 

According to the shell model of \cite{verhamme06}, this relation can be explained   if the galaxies with small Ly$\alpha$ shifts also have low HI column densities: indeed looking at their morphology, these galaxies also show compact  $\mathrm{Ly\alpha_{ext}/UV_{ext}}$, in agreement with the expectations. Compact Ly$\alpha$ spatial profile with respect to the continuum are indeed expected for galaxies with small N(HI), because the Ly$\alpha$ photons did not undergo many scatters with the neutral gas while escaping the galaxies (Verhamme et al. in preparation).  On the other hand, objects with no ISM outflows show large Ly$\alpha$ shifts ($\sim 500$ km/s). In the context of the shell model this should be observed only  for galaxies which have  large HI column densities ($\sim 10^{20}-10^{21} cm^{-2}$). Indeed these galaxies  also show large  $\mathrm{Ly\alpha_{ext}/UV_{ext}}$, in agreement with the expectations since in these cases the Ly$\alpha$ photons scatter many times before escaping, resulting in broader spatial profiles.

We finally do not observe any correlation between the trend observed for the IS and Ly$\alpha$ shifts (and therefore N(HI)) and any other physical properties such as  EW(Ly$\alpha$), SFR, Mass, sSFR, $\mathrm{\Sigma_{SFR}}$ and dust content.

    To better understand these correlations and assess their validity, we are  planning to enlarge the analysis carried out so far  to the entire VANDELS data-set, which will be available in the near future, since the present  work was applied only to half of the final VANDELS sample.  
    
    We finally remark that  the sources presented in this work have been selected as LBGs and not as LAEs, being therefore on average brighter in the UV continuum than the objects selected with a narrow band technique that are tuned to the Ly$\alpha$. This means that we are not probing the least massive sources, that are thought to be the most common galaxies in the very early Universe and possibly also those that contributed mostly to the reionization photon budget.  The correlations that we find could therefore be  stronger (or different) if investigated in a samples which could also include these lower mass galaxies. We are planning to test this hypothesis, by collecting a sample of strong LAEs, selected based on narrow band studies, to investigate how they compare with the VANDELS galaxies in the present sample and  if there are significant differences in the correlations that we observe based on the samples selection.

\begin{acknowledgements}
We thank the ESO staff for their continuous support for the VANDELS
survey, particularly the Paranal staff, who helped us to conduct the observations,
and the ESO user support group in Garching. FM and LP aknowledge financial support from Premiale 2015 MITiC. We would like to thank Alice Shapley and Max Pettini for their valuable comments on an earlier version of this manuscript.

\end{acknowledgements}

\bibliographystyle{aa} 
\bibliography{biblio}

\begin{thebibliography}{69}
\expandafter\ifx\csname natexlab\endcsname\relax\def\natexlab#1{#1}\fi

\bibitem[{{Amor{\'{\i}}n} {et~al.}(2017){Amor{\'{\i}}n}, {Fontana},
  {P{\'e}rez-Montero}, {Castellano}, {Guaita}, {Grazian}, {Le F{\`e}vre},
  {Ribeiro}, {Schaerer}, {Tasca}, {Thomas}, {Bardelli}, {Cassar{\`a}},
  {Cassata}, {Cimatti}, {Contini}, {de Barros}, {Garilli}, {Giavalisco},
  {Hathi}, {Koekemoer}, {Le Brun}, {Lemaux}, {Maccagni}, {Pentericci}, {Pforr},
  {Talia}, {Tresse}, {Vanzella}, {Vergani}, {Zamorani}, {Zucca}, \&
  {Merlin}}]{amorin17}
{Amor{\'{\i}}n}, R., {Fontana}, A., {P{\'e}rez-Montero}, E., {et~al.} 2017,
  Nature Astronomy, 1, 0052

\bibitem[{{Bottini} {et~al.}(2005){Bottini}, {Garilli}, {Maccagni}, {Tresse},
  {Le Brun}, {Le F{\`e}vre}, {Picat}, {Scaramella}, {Scodeggio}, {Vettolani},
  {Zanichelli}, {Adami}, {Arnaboldi}, {Arnouts}, {Bardelli}, {Bolzonella},
  {Cappi}, {Charlot}, {Ciliegi}, {Contini}, {Foucaud}, {Franzetti}, {Guzzo},
  {Ilbert}, {Iovino}, {McCracken}, {Marano}, {Marinoni}, {Mathez}, {Mazure},
  {Meneux}, {Merighi}, {Paltani}, {Pollo}, {Pozzetti}, {Radovich}, {Zamorani},
  \& {Zucca}}]{bottini05}
{Bottini}, D., {Garilli}, B., {Maccagni}, D., {et~al.} 2005, \pasp, 117, 996

\bibitem[{{Calzetti} {et~al.}(2000){Calzetti}, {Armus}, {Bohlin}, {Kinney},
  {Koornneef}, \& {Storchi-Bergmann}}]{calzetti00}
{Calzetti}, D., {Armus}, L., {Bohlin}, R.~C., {et~al.} 2000, \apj, 533, 682

\bibitem[{{Carnall} {et~al.}(2018){Carnall}, {Leja}, {Johnson}, {McLure},
  {Dunlop}, \& {Conroy}}]{carnall18}
{Carnall}, A.~C., {Leja}, J., {Johnson}, B.~D., {et~al.} 2018, arXiv e-prints
  [\eprint[arXiv]{1811.03635}]

\bibitem[{{Cassar{\`a}} {et~al.}(2016){Cassar{\`a}}, {Maccagni}, {Garilli},
  {Scodeggio}, {Thomas}, {Le F{\`e}vre}, {Zamorani}, {Schaerer}, {Lemaux},
  {Cassata}, {Le Brun}, {Pentericci}, {Tasca}, {Vanzella}, {Zucca},
  {Amor{\'{\i}}n}, {Bardelli}, {Castellano}, {Cimatti}, {Cucciati}, {Durkalec},
  {Fontana}, {Giavalisco}, {Grazian}, {Hathi}, {Ilbert}, {Paltani}, {Ribeiro},
  {Sommariva}, {Talia}, {Tresse}, {Vergani}, {Capak}, {Charlot}, {Contini}, {de
  la Torre}, {Dunlop}, {Fotopoulou}, {Guaita}, {Koekemoer},
  {L{\'o}pez-Sanjuan}, {Mellier}, {Pforr}, {Salvato}, {Scoville}, {Taniguchi},
  \& {Wang}}]{cassara16}
{Cassar{\`a}}, L.~P., {Maccagni}, D., {Garilli}, B., {et~al.} 2016, \aap, 593,
  A9

\bibitem[{{Cassata} {et~al.}(2013){Cassata}, {Le F{\`e}vre}, {Charlot},
  {Contini}, {Cucciati}, {Garilli}, {Zamorani}, {Adami}, {Bardelli}, {Le Brun},
  {Lemaux}, {Maccagni}, {Pollo}, {Pozzetti}, {Tresse}, {Vergani}, {Zanichelli},
  \& {Zucca}}]{cassata13}
{Cassata}, P., {Le F{\`e}vre}, O., {Charlot}, S., {et~al.} 2013, \aap, 556, A68

\bibitem[{{Castellano} {et~al.}(2012){Castellano}, {Fontana}, {Grazian},
  {Pentericci}, {Santini}, {Koekemoer}, {Cristiani}, {Galametz}, {Gallerani},
  {Vanzella}, {Boutsia}, {Gallozzi}, {Giallongo}, {Maiolino}, {Menci}, \&
  {Paris}}]{castellano12}
{Castellano}, M., {Fontana}, A., {Grazian}, A., {et~al.} 2012, \aap, 540, A39

\bibitem[{{Chabrier}(2003)}]{chabrier03}
{Chabrier}, G. 2003, \pasp, 115, 763

\bibitem[{{Chevallard} \& {Charlot}(2016)}]{beagle16}
{Chevallard}, J. \& {Charlot}, S. 2016, \mnras, 462, 1415

\bibitem[{{Cowie} {et~al.}(2011){Cowie}, {Barger}, \& {Hu}}]{cowie11}
{Cowie}, L.~L., {Barger}, A.~J., \& {Hu}, E.~M. 2011, \apj, 738, 136

\bibitem[{{Cullen} {et~al.}(2018){Cullen}, {McLure}, {Khochfar}, {Dunlop},
  {Dalla Vecchia}, {Carnall}, {Bourne}, {Castellano}, {Cimatti}, {Cirasuolo},
  {Elbaz}, {Fynbo}, {Garilli}, {Koekemoer}, {Marchi}, {Pentericci}, {Talia}, \&
  {Zamorani}}]{cullen18}
{Cullen}, F., {McLure}, R.~J., {Khochfar}, S., {et~al.} 2018, \mnras, 476, 3218

\bibitem[{{Dekel} {et~al.}(2009){Dekel}, {Birnboim}, {Engel}, {Freundlich},
  {Goerdt}, {Mumcuoglu}, {Neistein}, {Pichon}, {Teyssier}, \&
  {Zinger}}]{dekel09}
{Dekel}, A., {Birnboim}, Y., {Engel}, G., {et~al.} 2009, \nat, 457, 451

\bibitem[{{Du} {et~al.}(2018){Du}, {Shapley}, {Reddy}, {Jones}, {Stark},
  {Steidel}, {Strom}, {Rudie}, {Erb}, {Ellis}, \& {Pettini}}]{du18}
{Du}, X., {Shapley}, A.~E., {Reddy}, N.~A., {et~al.} 2018, \apj, 860, 75

\bibitem[{{Ebbets}(1995)}]{ebbets95}
{Ebbets}, D. 1995, in Calibrating Hubble Space Telescope. Post Servicing
  Mission, ed. A.~P. {Koratkar} \& C.~{Leitherer}, 207

\bibitem[{{Erb} {et~al.}(2016){Erb}, {Pettini}, {Steidel}, {Strom}, {Rudie},
  {Trainor}, {Shapley}, \& {Reddy}}]{erb16}
{Erb}, D.~K., {Pettini}, M., {Steidel}, C.~C., {et~al.} 2016, \apj, 830, 52

\bibitem[{{Finkelstein} {et~al.}(2015){Finkelstein}, {Finkelstein}, {Tilvi},
  {Malhotra}, {Rhoads}, {Grogin}, {Pirzkal}, {Dey}, {Jannuzi}, {Mobasher},
  {Pakzad}, {Salmon}, \& {Wang}}]{finkelstein15}
{Finkelstein}, K.~D., {Finkelstein}, S.~L., {Tilvi}, V., {et~al.} 2015, \apj,
  813, 78

\bibitem[{{Finkelstein} {et~al.}(2009){Finkelstein}, {Rhoads}, {Malhotra}, \&
  {Grogin}}]{finkelstein09}
{Finkelstein}, S.~L., {Rhoads}, J.~E., {Malhotra}, S., \& {Grogin}, N. 2009,
  \apj, 691, 465

\bibitem[{{Finkelstein} {et~al.}(2008){Finkelstein}, {Rhoads}, {Malhotra},
  {Grogin}, \& {Wang}}]{finkelstein08}
{Finkelstein}, S.~L., {Rhoads}, J.~E., {Malhotra}, S., {Grogin}, N., \& {Wang},
  J. 2008, \apj, 678, 655

\bibitem[{{Finkelstein} {et~al.}(2007){Finkelstein}, {Rhoads}, {Malhotra},
  {Pirzkal}, \& {Wang}}]{finkelstein07}
{Finkelstein}, S.~L., {Rhoads}, J.~E., {Malhotra}, S., {Pirzkal}, N., \&
  {Wang}, J. 2007, \apj, 660, 1023

\bibitem[{{Galametz} {et~al.}(2013){Galametz}, {Grazian}, {Fontana},
  {Ferguson}, {Ashby}, {Barro}, {Castellano}, {Dahlen}, {Donley}, {Faber},
  {Grogin}, {Guo}, {Huang}, {Kocevski}, {Koekemoer}, {Lee}, {McGrath}, {Peth},
  {Willner}, {Almaini}, {Cooper}, {Cooray}, {Conselice}, {Dickinson}, {Dunlop},
  {Fazio}, {Foucaud}, {Gardner}, {Giavalisco}, {Hathi}, {Hartley}, {Koo},
  {Lai}, {de Mello}, {McLure}, {Lucas}, {Paris}, {Pentericci}, {Santini},
  {Simpson}, {Sommariva}, {Targett}, {Weiner}, {Wuyts}, \& {the CANDELS
  Team}}]{galametz+13}
{Galametz}, A., {Grazian}, A., {Fontana}, A., {et~al.} 2013, \apjs, 206, 10

\bibitem[{{Garilli} {et~al.}(2010){Garilli}, {Fumana}, {Franzetti}, {Paioro},
  {Scodeggio}, {Le F{\`e}vre}, {Paltani}, \& {Scaramella}}]{garilli10}
{Garilli}, B., {Fumana}, M., {Franzetti}, P., {et~al.} 2010, \pasp, 122, 827

\bibitem[{{Garilli} {et~al.}(2012){Garilli}, {Paioro}, {Scodeggio},
  {Franzetti}, {Fumana}, \& {Guzzo}}]{garilli12}
{Garilli}, B., {Paioro}, L., {Scodeggio}, M., {et~al.} 2012, \pasp, 124, 1232

\bibitem[{{Grogin} {et~al.}(2011){Grogin}, {Kocevski}, {Faber}, {Ferguson},
  {Koekemoer}, {Riess}, {Acquaviva}, {Alexander}, {Almaini}, {Ashby}, {Barden},
  {Bell}, {Bournaud}, {Brown}, {Caputi}, {Casertano}, {Cassata}, {Castellano},
  {Challis}, {Chary}, {Cheung}, {Cirasuolo}, {Conselice}, {Roshan Cooray},
  {Croton}, {Daddi}, {Dahlen}, {Dav{\'e}}, {de Mello}, {Dekel}, {Dickinson},
  {Dolch}, {Donley}, {Dunlop}, {Dutton}, {Elbaz}, {Fazio}, {Filippenko},
  {Finkelstein}, {Fontana}, {Gardner}, {Garnavich}, {Gawiser}, {Giavalisco},
  {Grazian}, {Guo}, {Hathi}, {H{\"a}ussler}, {Hopkins}, {Huang}, {Huang},
  {Jha}, {Kartaltepe}, {Kirshner}, {Koo}, {Lai}, {Lee}, {Li}, {Lotz}, {Lucas},
  {Madau}, {McCarthy}, {McGrath}, {McIntosh}, {McLure}, {Mobasher},
  {Moustakas}, {Mozena}, {Nandra}, {Newman}, {Niemi}, {Noeske}, {Papovich},
  {Pentericci}, {Pope}, {Primack}, {Rajan}, {Ravindranath}, {Reddy}, {Renzini},
  {Rix}, {Robaina}, {Rodney}, {Rosario}, {Rosati}, {Salimbeni}, {Scarlata},
  {Siana}, {Simard}, {Smidt}, {Somerville}, {Spinrad}, {Straughn}, {Strolger},
  {Telford}, {Teplitz}, {Trump}, {van der Wel}, {Villforth}, {Wechsler},
  {Weiner}, {Wiklind}, {Wild}, {Wilson}, {Wuyts}, {Yan}, \& {Yun}}]{grogin11}
{Grogin}, N.~A., {Kocevski}, D.~D., {Faber}, S.~M., {et~al.} 2011, \apjs, 197,
  35

\bibitem[{{Gronke} \& {Dijkstra}(2016)}]{gronke16}
{Gronke}, M. \& {Dijkstra}, M. 2016, \apj, 826, 14

\bibitem[{{Guaita} {et~al.}(2017){Guaita}, {Talia}, {Pentericci}, {Verhamme},
  {Cassata}, {Lemaux}, {Orlitova}, {Ribeiro}, {Schaerer}, {Zamorani},
  {Garilli}, {Le Brun}, {Le F{\`e}vre}, {Maccagni}, {Tasca}, {Thomas},
  {Vanzella}, {Zucca}, {Amorin}, {Bardelli}, {Castellano}, {Grazian}, {Hathi},
  {Koekemoer}, \& {Marchi}}]{guaita17}
{Guaita}, L., {Talia}, M., {Pentericci}, L., {et~al.} 2017, \aap, 606, A19

\bibitem[{{Guo} {et~al.}(2013){Guo}, {Ferguson}, {Giavalisco}, {Barro},
  {Willner}, {Ashby}, {Dahlen}, {Donley}, {Faber}, {Fontana}, {Galametz},
  {Grazian}, {Huang}, {Kocevski}, {Koekemoer}, {Koo}, {McGrath}, {Peth},
  {Salvato}, {Wuyts}, {Castellano}, {Cooray}, {Dickinson}, {Dunlop}, {Fazio},
  {Gardner}, {Gawiser}, {Grogin}, {Hathi}, {Hsu}, {Lee}, {Lucas}, {Mobasher},
  {Nandra}, {Newman}, \& {van der Wel}}]{guo+13}
{Guo}, Y., {Ferguson}, H.~C., {Giavalisco}, M., {et~al.} 2013, \apjs, 207, 24

\bibitem[{{Gutkin} {et~al.}(2016){Gutkin}, {Charlot}, \& {Bruzual}}]{gutkin16}
{Gutkin}, J., {Charlot}, S., \& {Bruzual}, G. 2016, \mnras, 462, 1757

\bibitem[{{Hathi} {et~al.}(2016){Hathi}, {Le F{\`e}vre}, {Ilbert}, {Cassata},
  {Tasca}, {Lemaux}, {Garilli}, {Le Brun}, {Maccagni}, {Pentericci}, {Thomas},
  {Vanzella}, {Zamorani}, {Zucca}, {Amor{\'{\i}}n}, {Bardelli}, {Cassar{\`a}},
  {Castellano}, {Cimatti}, {Cucciati}, {Durkalec}, {Fontana}, {Giavalisco},
  {Grazian}, {Guaita}, {Koekemoer}, {Paltani}, {Pforr}, {Ribeiro}, {Schaerer},
  {Scodeggio}, {Sommariva}, {Talia}, {Tresse}, {Vergani}, {Capak}, {Charlot},
  {Contini}, {Cuby}, {de la Torre}, {Dunlop}, {Fotopoulou},
  {L{\'o}pez-Sanjuan}, {Mellier}, {Salvato}, {Scoville}, {Taniguchi}, \&
  {Wang}}]{hathi16}
{Hathi}, N.~P., {Le F{\`e}vre}, O., {Ilbert}, O., {et~al.} 2016, \aap, 588, A26

\bibitem[{{Heckman} {et~al.}(2015){Heckman}, {Alexandroff}, {Borthakur},
  {Overzier}, \& {Leitherer}}]{heckman15}
{Heckman}, T.~M., {Alexandroff}, R.~M., {Borthakur}, S., {Overzier}, R., \&
  {Leitherer}, C. 2015, \apj, 809, 147

\bibitem[{{Heckman} {et~al.}(2011){Heckman}, {Borthakur}, {Overzier},
  {Kauffmann}, {Basu-Zych}, {Leitherer}, {Sembach}, {Martin}, {Rich},
  {Schiminovich}, \& {Seibert}}]{heckman11}
{Heckman}, T.~M., {Borthakur}, S., {Overzier}, R., {et~al.} 2011, \apj, 730, 5

\bibitem[{{Jones} {et~al.}(2012){Jones}, {Stark}, \& {Ellis}}]{jones12}
{Jones}, T., {Stark}, D.~P., \& {Ellis}, R.~S. 2012, \apj, 751, 51

\bibitem[{{Koekemoer} {et~al.}(2011){Koekemoer}, {Faber}, {Ferguson}, {Grogin},
  {Kocevski}, {Koo}, {Lai}, {Lotz}, {Lucas}, {McGrath}, {Ogaz}, {Rajan},
  {Riess}, {Rodney}, {Strolger}, {Casertano}, {Castellano}, {Dahlen},
  {Dickinson}, {Dolch}, {Fontana}, {Giavalisco}, {Grazian}, {Guo}, {Hathi},
  {Huang}, {van der Wel}, {Yan}, {Acquaviva}, {Alexander}, {Almaini}, {Ashby},
  {Barden}, {Bell}, {Bournaud}, {Brown}, {Caputi}, {Cassata}, {Challis},
  {Chary}, {Cheung}, {Cirasuolo}, {Conselice}, {Roshan Cooray}, {Croton},
  {Daddi}, {Dav{\'e}}, {de Mello}, {de Ravel}, {Dekel}, {Donley}, {Dunlop},
  {Dutton}, {Elbaz}, {Fazio}, {Filippenko}, {Finkelstein}, {Frazer}, {Gardner},
  {Garnavich}, {Gawiser}, {Gruetzbauch}, {Hartley}, {H{\"a}ussler},
  {Herrington}, {Hopkins}, {Huang}, {Jha}, {Johnson}, {Kartaltepe},
  {Khostovan}, {Kirshner}, {Lani}, {Lee}, {Li}, {Madau}, {McCarthy},
  {McIntosh}, {McLure}, {McPartland}, {Mobasher}, {Moreira}, {Mortlock},
  {Moustakas}, {Mozena}, {Nandra}, {Newman}, {Nielsen}, {Niemi}, {Noeske},
  {Papovich}, {Pentericci}, {Pope}, {Primack}, {Ravindranath}, {Reddy},
  {Renzini}, {Rix}, {Robaina}, {Rosario}, {Rosati}, {Salimbeni}, {Scarlata},
  {Siana}, {Simard}, {Smidt}, {Snyder}, {Somerville}, {Spinrad}, {Straughn},
  {Telford}, {Teplitz}, {Trump}, {Vargas}, {Villforth}, {Wagner}, {Wandro},
  {Wechsler}, {Weiner}, {Wiklind}, {Wild}, {Wilson}, {Wuyts}, \&
  {Yun}}]{koekemoer+11}
{Koekemoer}, A.~M., {Faber}, S.~M., {Ferguson}, H.~C., {et~al.} 2011, \apjs,
  197, 36

\bibitem[{{Kornei} {et~al.}(2010){Kornei}, {Shapley}, {Erb}, {Steidel},
  {Reddy}, {Pettini}, \& {Bogosavljevi{\'c}}}]{kornei10}
{Kornei}, K.~A., {Shapley}, A.~E., {Erb}, D.~K., {et~al.} 2010, \apj, 711, 693

\bibitem[{{Le F{\`e}vre} {et~al.}(2017){Le F{\`e}vre}, {Lemaux}, {Nakajima},
  {Schaerer}, {Talia}, {Zamorani}, {Cassata}, {Garilli}, {Maccagni},
  {Pentericci}, {Tasca}, {Zucca}, {Amorin}, {Bardelli}, {Cimatti},
  {Giavalisco}, {Guaita}, {Hathi}, {Marchi}, {Vanzella}, {Vergani}, \&
  {Dunlop}}]{lefevre17}
{Le F{\`e}vre}, O., {Lemaux}, B.~C., {Nakajima}, K., {et~al.} 2017, ArXiv
  e-prints [\eprint[arXiv]{1710.10715}]

\bibitem[{{Le F{\`e}vre} {et~al.}(2015){Le F{\`e}vre}, {Tasca}, {Cassata},
  {Garilli}, {Le Brun}, {Maccagni}, {Pentericci}, {Thomas}, {Vanzella},
  {Zamorani}, {Zucca}, {Amorin}, {Bardelli}, {Capak}, {Cassar{\`a}},
  {Castellano}, {Cimatti}, {Cuby}, {Cucciati}, {de la Torre}, {Durkalec},
  {Fontana}, {Giavalisco}, {Grazian}, {Hathi}, {Ilbert}, {Lemaux}, {Moreau},
  {Paltani}, {Ribeiro}, {Salvato}, {Schaerer}, {Scodeggio}, {Sommariva},
  {Talia}, {Taniguchi}, {Tresse}, {Vergani}, {Wang}, {Charlot}, {Contini},
  {Fotopoulou}, {L{\'o}pez-Sanjuan}, {Mellier}, \& {Scoville}}]{lefevreVIMOS}
{Le F{\`e}vre}, O., {Tasca}, L.~A.~M., {Cassata}, P., {et~al.} 2015, \aap, 576,
  A79

\bibitem[{{Leja} {et~al.}(2018){Leja}, {Carnall}, {Johnson}, {Conroy}, \&
  {Speagle}}]{leja18}
{Leja}, J., {Carnall}, A.~C., {Johnson}, B.~D., {Conroy}, C., \& {Speagle},
  J.~S. 2018, arXiv e-prints [\eprint[arXiv]{1811.03637}]

\bibitem[{{Lenz} \& {Ayres}(1992)}]{lenz92}
{Lenz}, D.~D. \& {Ayres}, T.~R. 1992, \pasp, 104, 1104

\bibitem[{{Marchi} {et~al.}(2018){Marchi}, {Pentericci}, {Guaita}, {Schaerer},
  {Verhamme}, {Castellano}, {Ribeiro}, {Garilli}, {F{\`e}vre}, {Amorin},
  {Bardelli}, {Cassata}, {Durkalec}, {Grazian}, {Hathi}, {Lemaux}, {Maccagni},
  {Vanzella}, \& {Zucca}}]{marchi18}
{Marchi}, F., {Pentericci}, L., {Guaita}, L., {et~al.} 2018, \aap, 614, A11

\bibitem[{{Maseda} {et~al.}(2017){Maseda}, {Brinchmann}, {Franx}, {Bacon},
  {Bouwens}, {Schmidt}, {Boogaard}, {Contini}, {Feltre}, {Inami},
  {Kollatschny}, {Marino}, {Richard}, {Verhamme}, \& {Wisotzki}}]{maseda17}
{Maseda}, M.~V., {Brinchmann}, J., {Franx}, M., {et~al.} 2017, \aap, 608, A4

\bibitem[{{Matthee} {et~al.}(2018){Matthee}, {Sobral}, {Gronke},
  {Paulino-Afonso}, {Stefanon}, \& {R{\"o}ttgering}}]{matthee18}
{Matthee}, J., {Sobral}, D., {Gronke}, M., {et~al.} 2018, \aap, 619, A136

\bibitem[{{McLure} {et~al.}(2018){McLure}, {Pentericci}, {Cimatti}, {Dunlop},
  {Elbaz}, {Fontana}, {Nandra}, {Amorin}, {Bolzonella}, {Bongiorno}, {Carnall},
  {Castellano}, {Cirasuolo}, {Cucciati}, {Cullen}, {De Barros}, {Finkelstein},
  {Fontanot}, {Franzetti}, {Fumana}, {Gargiulo}, {Garilli}, {Guaita},
  {Hartley}, {Iovino}, {Jarvis}, {Juneau}, {Karman}, {Maccagni}, {Marchi},
  {M{\'a}rmol-Queralt{\'o}}, {Pompei}, {Pozzetti}, {Scodeggio}, {Sommariva},
  {Talia}, {Almaini}, {Balestra}, {Bardelli}, {Bell}, {Bourne}, {Bowler},
  {Brusa}, {Buitrago}, {Caputi}, {Cassata}, {Charlot}, {Citro}, {Cresci},
  {Cristiani}, {Curtis-Lake}, {Dickinson}, {Fazio}, {Ferguson}, {Fiore},
  {Franco}, {Fynbo}, {Galametz}, {Georgakakis}, {Giavalisco}, {Grazian},
  {Hathi}, {Jung}, {Kim}, {Koekemoer}, {Khusanova}, {Le F{\`e}vre}, {Lotz},
  {Mannucci}, {Maltby}, {Matsuoka}, {McLeod}, {Mendez-Hernandez},
  {Mendez-Abreu}, {Mignoli}, {Moresco}, {Mortlock}, {Nonino}, {Pannella},
  {Papovich}, {Popesso}, {Rosario}, {Salvato}, {Santini}, {Schaerer},
  {Schreiber}, {Stark}, {Tasca}, {Thomas}, {Treu}, {Vanzella}, {Wild},
  {Williams}, {Zamorani}, \& {Zucca}}]{mclure18}
{McLure}, R.~J., {Pentericci}, L., {Cimatti}, A., {et~al.} 2018, \mnras, 479,
  25

\bibitem[{{Nakajima} {et~al.}(2018){Nakajima}, {Schaerer}, {Le F{\`e}vre},
  {Amor{\'{\i}}n}, {Talia}, {Lemaux}, {Tasca}, {Vanzella}, {Zamorani},
  {Bardelli}, {Grazian}, {Guaita}, {Hathi}, {Pentericci}, \&
  {Zucca}}]{nakajima18}
{Nakajima}, K., {Schaerer}, D., {Le F{\`e}vre}, O., {et~al.} 2018, \aap, 612,
  A94

\bibitem[{{Osterbrock} \& {Ferland}(2006)}]{CIIIdoublet}
{Osterbrock}, D.~E. \& {Ferland}, G.~J. 2006, {Astrophysics of gaseous nebulae
  and active galactic nuclei}

\bibitem[{{Pardy} {et~al.}(2014){Pardy}, {Cannon}, {{\"O}stlin}, {Hayes},
  {Rivera-Thorsen}, {Sandberg}, {Adamo}, {Freeland}, {Herenz}, {Guaita},
  {Kunth}, {Laursen}, {Mas-Hesse}, {Melinder}, {Orlitov{\'a}},
  {Ot{\'{\i}}-Floranes}, {Puschnig}, {Schaerer}, \& {Verhamme}}]{pardy14}
{Pardy}, S.~A., {Cannon}, J.~M., {{\"O}stlin}, G., {et~al.} 2014, \apj, 794,
  101

\bibitem[{{Pentericci} {et~al.}(2009){Pentericci}, {Grazian}, {Fontana},
  {Castellano}, {Giallongo}, {Salimbeni}, \& {Santini}}]{pentericci09}
{Pentericci}, L., {Grazian}, A., {Fontana}, A., {et~al.} 2009, \aap, 494, 553

\bibitem[{{Pentericci} {et~al.}(2007){Pentericci}, {Grazian}, {Fontana},
  {Salimbeni}, {Santini}, {de Santis}, {Gallozzi}, \&
  {Giallongo}}]{pentericci07}
{Pentericci}, L., {Grazian}, A., {Fontana}, A., {et~al.} 2007, \aap, 471, 433

\bibitem[{{Pentericci} {et~al.}(2010){Pentericci}, {Grazian}, {Scarlata},
  {Fontana}, {Castellano}, {Giallongo}, \& {Vanzella}}]{pentericci10}
{Pentericci}, L., {Grazian}, A., {Scarlata}, C., {et~al.} 2010, \aap, 514, A64

\bibitem[{{Pentericci} {et~al.}(2018){Pentericci}, {McLure}, {Garilli},
  {Cucciati}, {Franzetti}, {Iovino}, {Amorin}, {Bolzonella}, {Bongiorno},
  {Carnall}, {Castellano}, {Cimatti}, {Cirasuolo}, {Cullen}, {De Barros},
  {Dunlop}, {Elbaz}, {Finkelstein}, {Fontana}, {Fontanot}, {Fumana},
  {Gargiulo}, {Guaita}, {Hartley}, {Jarvis}, {Juneau}, {Karman}, {Maccagni},
  {Marchi}, {Marmol-Queralto}, {Nandra}, {Pompei}, {Pozzetti}, {Scodeggio},
  {Sommariva}, {Talia}, {Almaini}, {Balestra}, {Bardelli}, {Bell}, {Bourne},
  {Bowler}, {Brusa}, {Buitrago}, {Caputi}, {Cassata}, {Charlot}, {Citro},
  {Cresci}, {Cristiani}, {Curtis-Lake}, {Dickinson}, {Fazio}, {Ferguson},
  {Fiore}, {Franco}, {Fynbo}, {Galametz}, {Georgakakis}, {Giavalisco},
  {Grazian}, {Hathi}, {Jung}, {Kim}, {Koekemoer}, {Khusanova}, {Le F{\`e}vre},
  {Lotz}, {Mannucci}, {Maltby}, {Matsuoka}, {McLeod}, {Mendez-Hernandez},
  {Mendez-Abreu}, {Mignoli}, {Moresco}, {Mortlock}, {Nonino}, {Pannella},
  {Papovich}, {Popesso}, {Rosario}, {Salvato}, {Santini}, {Schaerer},
  {Schreiber}, {Stark}, {Tasca}, {Thomas}, {Treu}, {Vanzella}, {Wild},
  {Williams}, {Zamorani}, \& {Zucca}}]{pentericci18}
{Pentericci}, L., {McLure}, R.~J., {Garilli}, B., {et~al.} 2018, \aap, 616,
  A174

\bibitem[{{Quider} {et~al.}(2009){Quider}, {Pettini}, {Shapley}, \&
  {Steidel}}]{quider09}
{Quider}, A.~M., {Pettini}, M., {Shapley}, A.~E., \& {Steidel}, C.~C. 2009,
  \mnras, 398, 1263

\bibitem[{{Reddy} {et~al.}(2010){Reddy}, {Erb}, {Pettini}, {Steidel}, \&
  {Shapley}}]{reddy10}
{Reddy}, N.~A., {Erb}, D.~K., {Pettini}, M., {Steidel}, C.~C., \& {Shapley},
  A.~E. 2010, \apj, 712, 1070

\bibitem[{{Reddy} {et~al.}(2006){Reddy}, {Steidel}, {Fadda}, {Yan}, {Pettini},
  {Shapley}, {Erb}, \& {Adelberger}}]{reddy06}
{Reddy}, N.~A., {Steidel}, C.~C., {Fadda}, D., {et~al.} 2006, \apj, 644, 792

\bibitem[{{Rigby} {et~al.}(2015){Rigby}, {Bayliss}, {Gladders}, {Sharon},
  {Wuyts}, {Dahle}, {Johnson}, \& {Pe{\~n}a-Guerrero}}]{rigby15}
{Rigby}, J.~R., {Bayliss}, M.~B., {Gladders}, M.~D., {et~al.} 2015, \apjl, 814,
  L6

\bibitem[{{Santini} {et~al.}(2017){Santini}, {Fontana}, {Castellano}, {Di
  Criscienzo}, {Merlin}, {Amorin}, {Cullen}, {Daddi}, {Dickinson}, {Dunlop},
  {Grazian}, {Lamastra}, {McLure}, {Micha{\l}owski}, {Pentericci}, \&
  {Shu}}]{santini17}
{Santini}, P., {Fontana}, A., {Castellano}, M., {et~al.} 2017, \apj, 847, 76

\bibitem[{{Schaerer} {et~al.}(2019){Schaerer}, {Fragos}, \&
  {Izotov}}]{schaerer19}
{Schaerer}, D., {Fragos}, T., \& {Izotov}, Y.~I. 2019, \aap, 622, L10

\bibitem[{{Scodeggio} {et~al.}(2005){Scodeggio}, {Franzetti}, {Garilli},
  {Zanichelli}, {Paltani}, {Maccagni}, {Bottini}, {Le Brun}, {Contini},
  {Scaramella}, {Adami}, {Bardelli}, {Zucca}, {Tresse}, {Ilbert}, {Foucaud},
  {Iovino}, {Merighi}, {Zamorani}, {Gavignaud}, {Rizzo}, {McCracken}, {Le
  F{\`e}vre}, {Picat}, {Vettolani}, {Arnaboldi}, {Arnouts}, {Bolzonella},
  {Cappi}, {Charlot}, {Ciliegi}, {Guzzo}, {Marano}, {Marinoni}, {Mathez},
  {Mazure}, {Meneux}, {Pell{\`o}}, {Pollo}, {Pozzetti}, \&
  {Radovich}}]{scodeggio05}
{Scodeggio}, M., {Franzetti}, P., {Garilli}, B., {et~al.} 2005, \pasp, 117,
  1284

\bibitem[{{Shapley} {et~al.}(2003){Shapley}, {Steidel}, {Pettini}, \&
  {Adelberger}}]{shapley03}
{Shapley}, A.~E., {Steidel}, C.~C., {Pettini}, M., \& {Adelberger}, K.~L. 2003,
  \apj, 588, 65

\bibitem[{{Shimasaku} {et~al.}(2006){Shimasaku}, {Kashikawa}, {Doi}, {Ly},
  {Malkan}, {Matsuda}, {Ouchi}, {Hayashino}, {Iye}, {Motohara}, {Murayama},
  {Nagao}, {Ohta}, {Okamura}, {Sasaki}, {Shioya}, \& {Taniguchi}}]{shimasaku06}
{Shimasaku}, K., {Kashikawa}, N., {Doi}, M., {et~al.} 2006, \pasj, 58, 313

\bibitem[{{Sommariva} {et~al.}(2012){Sommariva}, {Mannucci}, {Cresci},
  {Maiolino}, {Marconi}, {Nagao}, {Baroni}, \& {Grazian}}]{sommariva12}
{Sommariva}, V., {Mannucci}, F., {Cresci}, G., {et~al.} 2012, \aap, 539, A136

\bibitem[{{Stark} {et~al.}(2017){Stark}, {Ellis}, {Charlot}, {Chevallard},
  {Tang}, {Belli}, {Zitrin}, {Mainali}, {Gutkin}, {Vidal-Garc{\'{\i}}a},
  {Bouwens}, \& {Oesch}}]{stark+17}
{Stark}, D.~P., {Ellis}, R.~S., {Charlot}, S., {et~al.} 2017, \mnras, 464, 469

\bibitem[{{Stark} {et~al.}(2010){Stark}, {Ellis}, {Chiu}, {Ouchi}, \&
  {Bunker}}]{stark10}
{Stark}, D.~P., {Ellis}, R.~S., {Chiu}, K., {Ouchi}, M., \& {Bunker}, A. 2010,
  \mnras, 408, 1628

\bibitem[{{Stark} {et~al.}(2014){Stark}, {Richard}, {Siana}, {Charlot},
  {Freeman}, {Gutkin}, {Wofford}, {Robertson}, {Amanullah}, {Watson}, \&
  {Milvang-Jensen}}]{stark14}
{Stark}, D.~P., {Richard}, J., {Siana}, B., {et~al.} 2014, \mnras, 445, 3200

\bibitem[{{Steidel} {et~al.}(2018){Steidel}, {Bogosavlevic}, {Shapley},
  {Reddy}, {Rudie}, {Pettini}, {Trainor}, \& {Strom}}]{steidel18}
{Steidel}, C.~C., {Bogosavlevic}, M., {Shapley}, A.~E., {et~al.} 2018, ArXiv
  e-prints [\eprint[arXiv]{1805.06071}]

\bibitem[{{Talia} {et~al.}(2012){Talia}, {Mignoli}, {Cimatti}, {Kurk}, {Berta},
  {Bolzonella}, {Cassata}, {Daddi}, {Dickinson}, {Franceschini}, {Halliday},
  {Pozzetti}, {Renzini}, {Rodighiero}, {Rosati}, \& {Zamorani}}]{talia12}
{Talia}, M., {Mignoli}, M., {Cimatti}, A., {et~al.} 2012, \aap, 539, A61

\bibitem[{{Verhamme} {et~al.}(2015){Verhamme}, {Orlitov{\'a}}, {Schaerer}, \&
  {Hayes}}]{verhamme15}
{Verhamme}, A., {Orlitov{\'a}}, I., {Schaerer}, D., \& {Hayes}, M. 2015, \aap,
  578, A7

\bibitem[{{Verhamme} {et~al.}(2017){Verhamme}, {Orlitov{\'a}}, {Schaerer},
  {Izotov}, {Worseck}, {Thuan}, \& {Guseva}}]{verhamme17}
{Verhamme}, A., {Orlitov{\'a}}, I., {Schaerer}, D., {et~al.} 2017, \aap, 597,
  A13

\bibitem[{{Verhamme} {et~al.}(2006){Verhamme}, {Schaerer}, \&
  {Maselli}}]{verhamme06}
{Verhamme}, A., {Schaerer}, D., \& {Maselli}, A. 2006, \aap, 460, 397

\bibitem[{{Weiner} {et~al.}(2009){Weiner}, {Coil}, {Prochaska}, {Newman},
  {Cooper}, {Bundy}, {Conselice}, {Dutton}, {Faber}, {Koo}, {Lotz}, {Rieke}, \&
  {Rubin}}]{weiner09}
{Weiner}, B.~J., {Coil}, A.~L., {Prochaska}, J.~X., {et~al.} 2009, \apj, 692,
  187

\bibitem[{{Yamada} {et~al.}(2012){Yamada}, {Matsuda}, {Kousai}, {Hayashino},
  {Morimoto}, \& {Umemura}}]{yamada12}
{Yamada}, T., {Matsuda}, Y., {Kousai}, K., {et~al.} 2012, \apj, 751, 29

\bibitem[{{Yang} {et~al.}(2017){Yang}, {Malhotra}, {Gronke}, {Rhoads},
  {Leitherer}, {Wofford}, {Jiang}, {Dijkstra}, {Tilvi}, \& {Wang}}]{yang+17}
{Yang}, H., {Malhotra}, S., {Gronke}, M., {et~al.} 2017, ArXiv e-prints
  [\eprint[arXiv]{1701.01857}]

\end{thebibliography}
 
\end{document}